\documentclass[12pt]{article}
\usepackage{mathtools,amssymb,amsthm}
\usepackage{graphicx}
\usepackage{hyperref}
\usepackage{verbatim}
\usepackage{color}
\usepackage{float}
\usepackage{subcaption}
\usepackage[section]{placeins}
\usepackage{authblk}

\edef\restoreparindent{\parindent=\the\parindent\relax}
\usepackage{parskip}
\restoreparindent
\usepackage[normalem]{ulem}
\usepackage{fullpage}

\graphicspath{{images/}}

\numberwithin{equation}{section}

\newcommand{\be}{\begin{equation}}
\newcommand{\ee}{\end{equation}}
\newcommand{\eq}[1]{\begin{align}#1\end{align}}

\newcommand{\bfig}{\begin{figure}}
\newcommand{\efig}{\end{figure}}

\newcommand{\cL}{\mathcal{L}}

\newcommand{\ms}{\mathbb{S}}

\newcommand{\wsj}{W_\text{SJ}}

\newcommand{\spac}{\Sigma}

\newcommand{\im}{\mathrm{Image}}

\newcommand{\cO}{\mathcal O}
\newcommand{\hW}{\widehat W}
\newcommand{\hD}{\widehat \Delta}
\newcommand{\htD}{\widehat {\Gamma}}
\newcommand{\hO}{\widehat \Omega}
\newcommand{\ssee}{\mathcal S}

\newcommand{\mx}{\mathrm{max}}

\newcommand{\hK}{\widehat{K}}

\newcommand{\bv}{ \Phi}
\newcommand{\bk}{\mathbf k}

\newcommand{\bpp}{\mathbf p}
\newcommand{\bap}{{\bar {\mathbf p}}}

\newcommand{\bx}{\mathbf x}
\newcommand{\bu}{ \Psi}
\newcommand{\br}{\mathbf r}

\newcommand{\cM}{\mathcal{M}}

\newcommand{\cH}{\mathcal{H}}
\newcommand{\cS}{\mathcal{S}}
\newcommand{\cC}{\mathcal{C}}

\newcommand{\sseecyl}{\ssee_{\text{cyl}}}
\newcommand{\ket}[1]{\left| #1 \right>}
\newcommand{\bra}[1]{\left< #1 \right|}
\newcommand{\hrho}{\hat{\rho}}

\newcommand{\diam}{\mathbb D} 

\newcommand{\mink}{\mathbb M}

\newcommand{\av}[1]{\langle #1 \rangle}
\newcommand{\rin}[2]{{\langle #1 , #2\rangle}_R}
\newcommand{\lt}[1]{\langle #1 \rangle}
\newcommand{\con}{\alpha}
\newcommand{\nin}{\not \in} 
\newcommand{\nn}{\nonumber}
\newcommand{\hR}{\widehat R} 
\newcommand{\pone}{\psi^{(1)}}
\newcommand{\ptwo}{\psi^{(2)}}
\newcommand{\tpone}{\varphi^{(1)}}
\newcommand{\tptwo}{\varphi^{(2)}}
\newcommand{\hp}{\hat{p}}
\newcommand{\hq}{\hat{q}}
\newcommand{\baa}{\mathbf a}
\newcommand{\bbb}{\mathbf b} 
\begin{document}

\title{Spacetime Entanglement Entropy: \\ Covariance and  Discreteness}

\author[1]{Abhishek Mathur} 
\author[1]{Sumati Surya\footnote{ssurya@rri.res.in}} 
\author[1,2]{Nomaan X}

\affil[1]{\small Raman Research Institute, Sadashivanagar, Bangalore 560080, India}
\affil[2]{\small Department of Mathematics and Statistics, University of New Brunswick, Fredericton, NB, Canada E3B 5A3}

\date{}
\maketitle
\begin{abstract}
We review some recent results on Sorkin's  spacetime formulation of the
entanglement entropy (SSEE)  
for a free quantum scalar field both in the continuum and in manifold-like causal sets.  The SSEE  for a causal diamond
in a  2d cylinder spacetime has been shown to have a  Calabrese-Cardy form, while for  de
Sitter and Schwarzschild de Sitter horizons in dimensions $d>2$,  it matches the mode-wise von-Neumann entropy.   In
these continuum examples the SSEE is regulated by imposing a UV cut-off.  Manifold-like causal sets come with a natural
covariant spacetime cut-off and thus provide an arena to study regulated QFT. However,  the SSEE for different manifold-like causal sets in $d=2$ and
$d=4$  has been shown to exhibit  a volume rather than an area law. The  area law is
recovered only when an additional UV cut-off is implemented in the scaling regime of the spectrum which  mimics the
continuum   behaviour. We discuss the implications of
these results and suggest that a volume-law may be a manifestation of the fundamental non-locality
of  causal sets and a sign of new UV physics.  

\end{abstract}

\section{Introduction}  

It is now well established that entanglement entropy (EE)   is a   useful tool for measuring  both  the entanglement
between subsystems as well as accounting for the Entropy-Area law in a diverse range of systems, including 
black holes and other spacetimes  with horizons \cite{bkls}.  
Unlike the classical entropy associated with a box of gas which is extensive, the EE is expected to satisfy complementarity, which in
turn 
implies that the entanglement is localised to the boundary separating the system
from its environment. More broadly, this is true of local QFTs with UV fixed points. While this may seem  to be a
general, and even defining picture of EE (thus linking it naturally to holography \cite{ryutag}), it is only a part of the story.
The EE for systems with long range  interactions or which are non-local do not necessarily satisfy an area law. In
\cite{elisa} the EE for subsystems in long-range Ising and 
Kitaev models were seen to follow either a volume law or an area law  depending on the exponent $\alpha$ in the fall off
$r^{-\alpha}$. Volume laws have also  been  shown for non-local QFTs like non-commutative
field theories as well as scalar  QFT  with non-local interactions on a lattice 
\cite{arXiv:1310.8345,takyanagi,urbanapaper}.   
These provide interesting counter-examples to many of  the standard discussions on  area laws and complementarity of the
von-Neumann entropy.  In particular, taking our cue from condensed matter systems where this has been related to
quantum phases \cite{elisa,urbanapaper,Nakagawa_2018} and the absence of conformal invariance, it seems pertinent to revisit our assumptions about
the nature of microscopic black hole area laws.

Since the finiteness of the EE in a QFT depends on the UV cut-off, a starting point would be to ask whether
there are hidden assumptions about the UV  behaviour of the QFT.  In the standard calculations it is assumed that a change in
the cut-off $l_c$ only serves to rescale the units  in which the area is measured, i.e. that {$\ln (S_{EE}) =
  -(d-2)\ln l_c  +b$}.
Given the
experience from condensed matter systems this is perhaps too strong an assumption for a theory of quantum gravity,
especially one in which non-locality might play an important role, as suggested in \cite{urbanapaper}.  

Without a complete theory of quantum gravity, this may seem hopelessly speculative. However, since the EE for a QFT is
calculated in a regime in which one still has a separation between the field and the background spacetime,  we can
study the EE of QFT on models of quantum gravity inspired spacetime, which are UV complete. 
Causal set theory provides us
with  one such concrete example, where the continuum spacetime is replaced by an ensemble of  randomly generated  locally finite posets or causal sets, with the order relation corresponding to  the spacetime causal relation. 
This picture of spacetime is motivated by the Hawking-King-McCarthy-Malament theorem, which says that the causal
structure poset  of  distinguishing spacetimes determines its conformal class \cite{blms,lr}.

The  random discretisation of spacetime provided  by causal sets theory is  thus  a useful arena in which to test these
ideas. Despite being discrete, causal sets do not violate local Lorentz invariance and provide a fundamental covariant
spacetime cut-off. The free scalar QFT on manifold like  causal sets was first studied in $d=2$ and $d=4$ Minkowski
spacetime starting from the Green's function \cite{johnston}. Because of the combination of discreteness and covariance 
in a causal set, Cauchy hypersurfaces are ill-defined and hence so are 
equal-time commutation relations. These can however be  replaced by the covariant Peierls bracket,
which as  shown by
Jonhston \cite{johnstontwo} provides a novel route  to quantisation  via whats now called the Sorkin-Johnston (SJ) vacuum
\cite{johnstonthesis,sorkinSJ,fv,Brumfred}.      

\begin{figure}[ht]
  \centering \resizebox{3.5in}{!}  {\includegraphics[width=\textwidth]{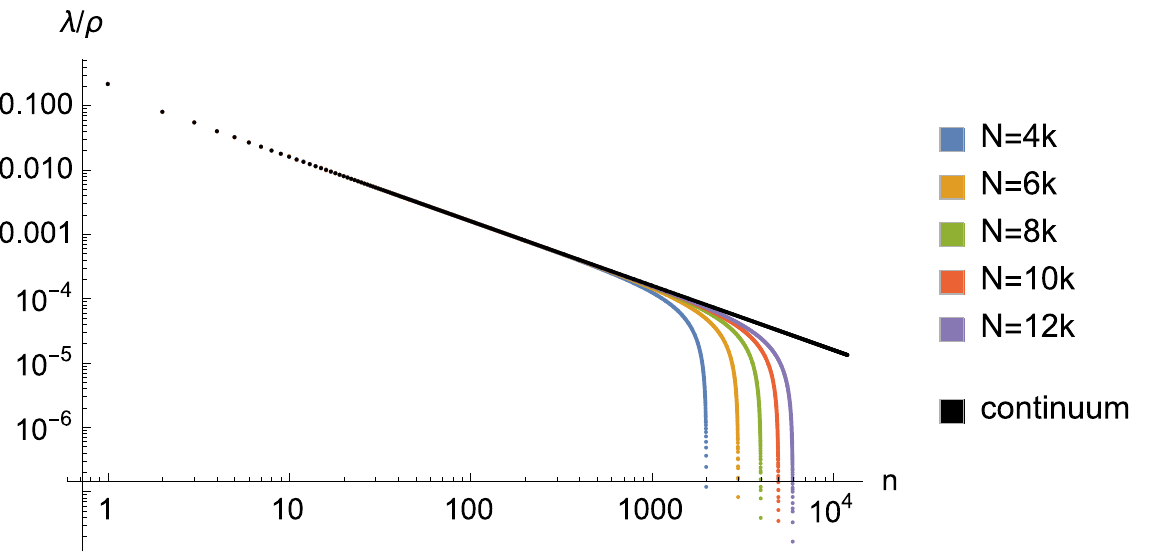}}
%\vspace{0.5cm}
\caption{{\small A log-log plot of the SJ spectrum wavelength $\lambda$ versus quantum number $n$. The black solid line
    is the continuum spectrum which exhibits a scale invariance. The coloured plots are the causal set spectrum for
    different discreteness scales which mimics the continuum scale invariance upto a ``knee'' beyond which it is explicitly
    broken. }}\label{SJspectrum.fig}
  \end{figure}

Since standard calculations of the EE also require Cauchy hypersurfaces, one needs a spacetime formulation of the EE for
causal  sets.   
Sorkin's spacetime entanglement entropy (SSEE) formulation for Gaussian free scalar fields provides an alternative 
measure for entanglement in terms of spacetime correlators \cite{ssee}.  In the continuum the SSEE has been calculated
for different types of horizons, in $d \geq 2$ and shown to satisfy the expected area law behaviour
\cite{ssy,Mathur:2021zzl,ourwork2}.  On a causal set the SSEE can be  calculated
using  the SJ vacuum, given the discrete Green's function. Sorkin and Yazdi first calculated the SSEE  
in causal sets approximated by a pair of nested $d=2$ Minkowski causal
diamonds \cite{soryaz} and more recently it has been calculated for $d=2,4$ de Sitter horizons as well as nested causal diamonds  in $d=4$ Minkowski
spacetime. In all cases, the SSEE for the causal set exhibits   a volume rather than an area law.  As shown by Sorkin
and Yazdi, the area law is
recovered only when an additional UV truncation is imposed on the SJ spectrum in the part of the spectrum that exhibits a 
continuum-like scaling-behaviour. Fig \ref{SJspectrum.fig} shows the comparison of the  continuum SJ spectrum with
the causal set spectrum for different discreteness scales. 
As shown in \cite{syx-entropy} the observations of \cite{soryaz} for the  $d=2$  nested causal diamonds  hold  more generally for
the de Sitter case as well. This  general behaviour in  all these cases can be related to some gross common  features in the
the causal set SJ  spectrum. For small quantum
number $n$ the spectrum has a 
continuum-like scaling regime $\lambda \propto n^{-\alpha}$ which  develops into a linear regime $\lambda \propto n$  in
the far UV.  
    
 {In Section \ref{SSEE.sec} we  review the construction of the SSEE \cite{ssee}. In Section \ref{continuum.sec} we discuss results
   in the the continuum which show that the  SSEE 
gives the expected area laws for $d>2$ de Sitter and Schwarzschild-de Sitter horizons \cite{ourwork2}. For the  $d=2$ causal diamond in
the cylinder spacetime the Calabrese-Cardy form for the EE is recovered, but the coefficients are not universal
\cite{Mathur:2021zzl}. In Section \ref{causalset} we summarise results on the 
causal set SSEE volume law as well as the truncation dependent area law \cite{soryaz,syx-entropy}. We analyse the SJ
spectrum and show that it follows a scaling-behaviour which then transitions to a linear behaviour in the UV. In Section
\ref{discussion} we end with a brief discussion of our results.}

\section{Sorkin's Spacetime Entanglement Entropy}
\label{SSEE.sec}

In a manifold-like  causal set non-locality implies that while there are analogs of spacelike hypersurfaces, these
cannot in any sense be Cauchy\footnote{See \cite{lr} and references therein for a review of causal set theory.}. In
order to define and study the EE of  QFT on causal sets therefore one needs  a  spacetime approach to quantisation and
additionally a spacetime formulation of the EE.   

 For a free Gaussian scalar field, and for a compact region of spacetime or equivalently for a finite causal  set, such 
 a spacetime quantisation is possible via the
 Sorkin-Johnston formulation.  Sorkin's spacetime Entanglement Entropy (SSEE) formula  in terms of
field  correlators then gives us the requisite form of the EE for causal sets.  We describe these constructions below. Note
that in all that follows we assume a  free Gaussian scalar 
field, and unless specified otherwise, compact spacetime regions.

The alternative to equal time commutation relations is the Peierel's bracket
\begin{equation}
[\phi(x), \phi(x')] = i \Delta (x,x') 
  \end{equation} 
  where the Pauli-Jordan operator $\Delta (x,x')= G_R(x,x')-G_A(x,x') $.  In \cite{johnstontwo}  this was used as a starting point for defining a free scalar field  Quantum Field Theory on a manifold-like causal set for which the advanced and retarded
  Green functions $G_{A,R}(x,x')$ are known. We describe this ``Sorkin-Johnston'' approach to defining the free scalar
  field Quantum Field Theory, first in the continuum and then subsequently in the causal set where it was first
  discovered \cite{johnston}.  

  For a compact globally  hyperbolic spacetime region $(M,g)$ the integral operator
  \begin{equation} 
i \hD \circ \psi(x) \equiv i \int dV' \Delta (x,x') \psi(x') 
    \end{equation} 
is self-adjoint. Since $\mathrm{ker} \Box = \mathrm{Im}  \Delta$, the eigenbasis of $i\hD$, which we call the
Sorkin-Johnston (SJ) basis,  provides a unique
and covariant mode decomposition for the quantum field.  The  spectral decomposition of $i\hD$ in terms of the SJ 
basis is  
\begin{equation}
i \Delta (x,x')= \sum_k \lambda_k u_k(x) u_k^*(x') - \sum_k \lambda_k u_k^*(x) u_k(x')
\end{equation}
where we have used the fact  that the SJ eigenvalues come in pairs $(\lambda_k, -\lambda_k)$ with corresponding
eigenfunctions $(u_k(x),u_k^*(x))$, $\lambda_k>0$. The novel insight in \cite{johnstontwo} was the recognition that this
suffices 
to define the Wightman function, and hence a covariant quantum vacuum, i.e. 
\begin{equation} 
\wsj(x,x') = \mathrm{Pos} (i \Delta) (x,x') = \sum_k \lambda_k u_k(x) u_k^*(x'). 
\end{equation}
This  SJ vacuum has been extensively  studied both in the continuum and in causal sets, where it  
was first obtained \cite{johnstontwo,johnstonthesis,sorkinSJ,aas,fv,Brumfred,ab,abdrsy,syx}.

In studying the EE of quantum fields, one typically looks at entanglement between a spatial region $U_\Sigma$  on a Cauchy
hypersurface $\Sigma$  and its complement $U_\Sigma^c$ in $\Sigma$. However, there is  a more natural underlying spacetime picture of 
entanglement, commonly used in AQFT, which is an entanglement between the spacetime regions $D(U_\Sigma)$
and  its causal complement  $D(U_\Sigma^c)$, where $D(X)$ denotes the domain of dependence of the region  $X$ in $(M,g)$. Such a spacetime
picture is particularly appealing for causal sets and  is the starting point for the construction of the SSEE for a
Gaussian scalar field.

Consider a globally hyperbolic spacetime $(M,g)$ and let $W(x,x')$ be the vacuum Wightman function associated with a
free scalar field in $(M,g)$. This can be expressed as  
\begin{equation}
  W(x,x') = R_0(x,x') + \frac{i}{2} \Delta(x,x') 
\end{equation}
where $R_0(x,x') = \sqrt{-\Delta^2}(x,x')$  is real,  symmetric  and positive semi-definite (this follows from the
positivity of $W(x,x')$). Restricting $W(x,x')$ to a globally hyperbolic compact sub region $\cO$
typically gives rise to a mixed state 
\begin{equation} 
 W(x,x') = R(x,x') + \frac{i}{2} \Delta(x,x') 
\end{equation} 
where $R(x,x')$ is no longer related to the Pauli-Jordan matrix but is still real, symmetric and positive definite.

We now sketch the construction of the SSEE \cite{ssee} in the continuum for a compact region $\cO \subset M$ in a spacetime
$(M,g)$.  We  define integral operators in $\cO$ via their integral kernels  and the associated $\cL^2$ inner product: 
\begin{eqnarray} 
\hat A \circ f (x)  &\equiv&  \int dV' A(x,x') f(x'), \nn \\
  \langle f, \hat  A \circ g\rangle &=& \int dV f^*(x) \hat A \circ g (x)
= \int dV dV' f^*(x) A (x,x') g(x'). 
\end{eqnarray}
$\hW$  is postive semi-definite and self-adjoint  with respect to the $\cL^2$ norm. Hence $\ker(\hR) \subseteq \ker(\hD)$ or equivalently $\im(\hR) \supseteq \im(\hD)$.% \mnote{Is it true that they are equal?} 

In what follows we  restrict to $\im(\hR)$.  Notice that  $\hR$ is symmetric and  positive definite for $f \in \im(\hR)$ and can therefore be viewed as a ``metric'', with a  symmetric inverse 
\begin{equation}
(\hR^{-1} \circ \hR) (x,x')  = (\hR\circ \hR^{-1})(x,x') = \delta(x-x').  \label{Rinv.eq}
\end{equation}
We can use it to  define the new operators 
\begin{eqnarray} 
  i \htD &\equiv&   i \hR^{-1}\circ \hD \Rightarrow  i \hD= i \hR \circ \htD  \nn \\
 \hO &\equiv& - \htD \circ \htD,  
\end{eqnarray}
so that  $\htD$  can be thought of as  $\hD$ with one ``lowered index''. 
Since $\hD \circ  f \in \im(\hD) \subseteq  \im(\hR) $, $\htD$ is well defined and  $\ker \htD=\ker \hD$. The operator
$\hO$ is moreover  positive semi-definite with respect to the $\hR$ norm
\begin{equation}
\rin{f}{g} = \int dVdV'f^*(x) R(x,x') g(x'). 
  \end{equation} 
since 
\begin{equation} 
\rin{ f}{  \hO \circ f}   = \rin{f}{ i \htD \circ i \htD f } = \rin{ i \htD \circ f}{  i \htD
 \circ f} \geq 0  
\end{equation}
for  $f \in \im(\hR)$. Thus the eigenvalues $\{\sigma_k^2 \}$  of $\hO$ are all positive.  They are moreover degenerate
since the eigenfunctions come in pairs $(\psi, \psi^*)$, which in turn means that we can use a  real eigenbasis $\{\pone_k,
\ptwo_k\}$, where  $\ptwo_k= \sigma_k^{-1} \htD \circ  \pone_k$, satisfying
\begin{equation}
\rin{\pone}{\pone} = \rin{\ptwo}{\ptwo}=1,  \quad \rin{\pone}{\ptwo}=0
  \end{equation} 
  Since $\hO$  is symmetric and therefore self adjoint, it admits a  
spectral decomposition 
\begin{equation}
  \Omega (x,x') = \sum_k  \sigma_k^2\biggl(  \pone_k(x) \tpone_k(x')+  \ptwo_k(x) \tptwo_k(x')\biggr)
  \end{equation} 
where $\tpone_k(x) \equiv (\hR \circ \pone_k)(x), \tptwo_k(x) \equiv (\hR \circ \ptwo_k)(x) $. Hence 
\begin{equation}
  i\htD (x,x') = \sum_k  \sigma_k\biggl( \pone_k(x) \tptwo_k(x') - \ptwo_k(x) \tpone_k(x')  \biggr),  
  \end{equation} 
with the related two point function $\hK = \hR^{-1}\circ \hW = {\mathbf 1} + \frac{i}{2} \htD$,  having  the ``block'' diagonal form
\eq{ K(x,x') \! = \!\sum_k   \pone_k(x) \tpone_k(x') +  \ptwo_k(x) \tptwo_k(x')  +\frac{i}{2} \sigma_k\biggl( \pone_k(x)
  \tptwo_k(x') - \ptwo_k(x) \tpone_k(x')\biggr).  \label{blockdiagK.eq}}
Since each block is  decoupled from all
others it can be viewed as a single pair of harmonic oscillators whose entropy can then be calculated relatively easily.

Since $\im(\hO)=\im(\hD)= \ker \Box$, the eigenfunctions $\{\pone_k, \ptwo_k \}$  of $\hO$ can be used for  a mode
decomposition for the quantum field. 
In  the harmonic oscillator basis,
\eq{\hat \Phi (x) = \sum_k \biggl(\hq_k \tpone_k(x) + \hp_k \tptwo_k(x) \biggr), \quad
  [\hq_k,\hp_{k'}]=i \sigma_k\delta_{k,k'}.\label{hphq.eq}}
which is consistent with the commutator $i \Delta(x,x') = [\hat \Phi(x), \hat \Phi(x')]$. In this basis 
\begin{eqnarray} 
K(x,x') & =&  < (\hR^{-1} \circ \hat \Phi(x))  \hat \Phi(x') >  \nn \\ &=&  \sum_{k} \biggl( <\hq_k \hq_{k}> 
             \pone_k(x)  \tpone_k(x') +  
  < \hq_k \hp_{k}> 
  \pone_k(x) \tptwo_k(x') \nn \\
  && + <\hp_k \hq_{k}>  
\ptwo_k(x) \tpone_k(x')+  <\hp_k \hp_{k}> 
\ptwo_k(x)  \tptwo_k(x') \biggr). 
\end{eqnarray}
where the block diagonalisation is obtained by comparing with $\mathbf 1=\frac{1}{2} (\hK + \hK^*)$ and $\htD=\frac{1}{2 i} (\hK
- \hK^*)$.  For each $k$ therefore we  have a harmonic oscillator with 
\begin{equation} 
  \hK_k \equiv
  \begin{pmatrix}
    <\hq_k \hq_k> &   <\hq_k \hp_k>\\
    <\hp_k \hq_k>& <\hp_k \hp_k>
  \end{pmatrix}  =  \begin{pmatrix}
   1 &  \frac{ i}{2} \sigma_k\\
   -\frac{ i}{2} \sigma_k & 1 
  \end{pmatrix}, \quad  i\htD_k \equiv \begin{pmatrix}
   0 &  i \sigma_k\\
   -i \sigma_k & 0  
 \end{pmatrix},
 \label{singleharm.eq}
\end{equation}
The problem thus reduces to a single degree  of freedom  with $\hK_k$ representing the two-point correlation functions
for a single harmonic oscillator in a given state, so that we can from now on drop  the index $k$.

{ Associated with any Gaussian state}  is a density matrix of the form 
\begin{equation}
\rho(q,q') = \sqrt{\frac{A}{\pi}} e^{-\frac{A}{2} (q^2 +q'^2) + i \frac{B}{2} (q^2-q'^2)-\frac{C}{2} (q-q')^2}.\label{gdm.eq}
  \end{equation} 
  One can use the replica trick \cite{Chen:2020ild,Keseman:2021dkf} to find the von Neumann entropy of $\rho$ to be  
  \begin{equation}
    S= - \frac{ \mu \ln \mu + (1-\mu) \ln (1-\mu) }{1-\mu},\label{rts.eq}
  \end{equation}
  where $\mu=\frac{\sqrt{1+2C/A} -1}{\sqrt{1+2C/A} + 1}$. The task at hand is to find the density matrix
  associated with the state Eqn. \eqref{singleharm.eq}. 

Consider the position  eigenbasis $\{\ket{q}\}$ of $\hq$, in which $\hq\ket{q} = q\ket{q}$, $\hp\ket{q}=-i\sigma\partial_q\ket{q}.$
The correlators $<\hat\eta_a \hat \eta_b > = \mathrm{Tr}(\hat \eta_a \hat \eta_b \hrho)$, for $(\hat \eta_1,
\hat \eta_2) \equiv (\hq,\hp)$ can then be explicitly calculated in this basis, using $\bra{q} \hrho \ket{q'} =\rho(q,q')$ as in Eqn
\eqref{gdm.eq}.  The integrals reduce to  the Gaussians $\int dq \, q^2\,  e^{-Aq^2}$ and are easily evaluated to
give 
\eq{<\hq\hq> = 1/(2A), \quad 
<\hq\hp> = \frac{ i}{2}\sigma,  \quad, 
<\hp\hq> =-\frac{ i}{2}\sigma,  \quad, 
<\hp\hp> = \sigma^2(A/2+C), }
where we have used the fact that $<qp>$ is purely imaginary and $<qq>$ and $ <pp>$ are purely real from Eqn
\eqref{singleharm.eq} to put $B=0$.

Equating to Eqn. \eqref{singleharm.eq} gives  $A=1/2$ and $C=1/\sigma^2 - 1/4$, so that $\mu=\frac{2-\sigma}{2+\sigma}$
for the associated single oscillator  von Neumann entropy Eqn. \eqref{rts.eq}.  
Following the work of \cite{Chen:2020ild}, we notice that  the eigenvalues of $(i\htD)^{-1}\hK$  are 
\eq{\mu^\pm = \frac{1}{2}\pm\sqrt{\frac{1}{4}+\frac{C}{2A}}=\frac{1}{2} \pm \frac{1}{\sigma},}
in terms of which Eqn.~\eqref{rts.eq} simplifies to 
\eq{S= \mu^+\ln|\mu^+| + \mu^-\ln|\mu^-|.}
Since the eigenvalues of $(i\htD)^{-1}\hK$ are same as the generalised eigenvalues of $\hW\circ\Psi = i\mu\hD\circ\Psi$,
we come to the SSEE formula after summing over all $k$: 
\begin{equation}
\hW \circ \Psi_k(x)  = i  \mu  \hD \Psi_k(x), \quad \Psi_k(x) \nin \mathrm{ker} (i \hD), \quad
\mathcal S= \sum_{\mu} \mu \ln |\mu|.
\label{ssee.eq} 
\end{equation} 

As formulated, the SSEE does not require us to use the SJ two point function and can therefore be used more widely  as
we will see in the following section. On the causal set this formulation has the clear advantage that one only 
requires $i\hD$ and $\hW$ to calculate the SSEE of a subcausal set and its causal complement. In turn, these
operators are well defined on the causal set via the SJ prescription.

We now review the SSEE construction in the continuum before moving on to the causal set. 

\section{SSEE in the Continuum}
\label{continuum.sec}

As discussed in the introduction, we are interested in finding the EE for a free scalar field for a globally hyperbolic
region  $\cO $ in a spacetime $(\cM,g)$, $\cO \subset \cM$ with respect to its causal complement. In the SSEE Eqn.~\eqref{ssee.eq} the mixed state $\hW|_{\cO}$ is obtained by restricting the pure (vacuum) state $\hW$ in $(\cM,g)$  to the
region $\cO$.   As an integral kernel, of course $W(x,x')|_\cO = W(x,x')$, but as an {\sl integral operator}  $\hW_\cO$ is distinct, since it only operates in the region $\cO$.   

In order to simplify the generalised  eigenvalue equation needs  we use a  mode decomposition with respect to the  two
sets of modes  $\{ \bv_{\bk}\}$ and $\{\bu_{\bpp}\} $ in $\cM$ and $\cO$, respectively. In general these modes are
typically required to be KG orthonormal 
\begin{eqnarray} 
(\bv_{\bk},\bv_{\bk'})_{\cM} = - (\bv_{\bk}^*,\bv_{\bk'}^*)_{\cM} & = & 
  \delta_{\bk\bk'}\;\;\text{and}\,\,(\bv_{\bk},\bv_{\bk'}^*)_{\cM}=0, \nn \\ 
(\bu_{\bpp},\bu_{\bpp'})_{\cM} = - (\bu_{\bpp}^*,\bu_{\bpp'}^*)_{\cM} & = & 
                                                                            \delta_{\bpp\bpp'}\;\;\text{and}\,\,(\bu_{\bpp},\bu_{\bpp'}^*)_{\cM}=0,
                                                                            \label{kgortho.eq}       
\end{eqnarray} 
where the KG inner product is
\begin{equation}
  (\phi_1,\phi_2)_{\cM} =  i\int_{\spac_{\cM}} d\spac^a \; \left(\phi_1^*\partial_a\phi_2 -
                          \phi_2\partial_a\phi_1^*\right), \label{kgip.eq}
\end{equation} 
with   $d\spac^a$ being  the volume element on a Cauchy hypersurface $\spac\in\cM$ with respect to the future pointing unit
normal.

Starting  with the vacuum state $\hW$ in  $(M,g)$ its restriction to $\hW|_\cO$ can be expressed in terms of the modes $\{\bu_{\bpp}\} $ which provide a complete basis in $\cO$
\begin{eqnarray} 
  W(\bx,\bx')\Big|_\cO  & = \sum_\bk \bv_{\bk}(\bx)\bv_{\bk}^*(\bx') &= \sum_{\bpp\bpp'}\Big(A_{\bpp\bpp'}\bu_{\bpp}(\bx)\bu_{\bpp'}^*(\bx')+
       B_{\bpp\bpp'}\bu_{\bpp}(\bx)\bu_{\bpp'}(\bx')\nn\\&& + C_{\bpp\bpp'}\bu_{\bpp}^*(\bx)\bu_{\bpp'}^*(\bx') + D_{\bpp\bpp'}\bu_{\bpp}^*(\bx)\bu_{\bpp'}(\bx')\Big),\label{wo.eq}
\end{eqnarray}
where
\begin{equation} 
  \bv_\bk(\bx)\Big|_\cO =  \sum_\bpp \left(\alpha_{\bk\bpp}\bu_\bpp(\bx) + \beta_{\bk\bpp}\bu_\bpp^*(\bx)\right)
\end{equation} 
with $\alpha_{\bk\bpp} = (\bu_\bpp,\bv_\bk)_\cO$ and $\beta_{\bk\bpp} = -(\bu_\bpp^*,\bv_\bk)_\cO$ and 
\begin{equation} 
A_{\bpp\bpp'} \equiv \sum_{\bk}\alpha_{\bk\bpp}\alpha_{\bk\bpp'}^*,\;\;B_{\bpp\bpp'} \equiv \sum_{\bk}\alpha_{\bk\bpp}\beta_{\bk\bpp'}^*,\;\; C_{\bpp\bpp'} \equiv \sum_{\bk}\beta_{\bk\bpp}\alpha_{\bk\bpp'}^*,\;\; D_{\bpp\bpp'} \equiv \sum_{\bk}\beta_{\bk\bpp}\beta_{\bk\bpp'}^*.\label{abcd.eq}
\end{equation}
Expanding the Pauli-Jordan function $i\Delta(\bx,\bx')=[\hat \Phi(\bx), \hat \Phi(\bx')]$  in the $\{\bu_{\bpp}\} $
modes 
\eq{
i\Delta(\bx,\bx')=\sum_\bpp \left(\bu_{\bpp}(\bx)\bu_{\bpp}^*(\bx') - \bu_{\bpp}^*(\bx)\bu_{\bpp}(\bx')\right).\label{ido.eq}
}
the generalised eigenvalue equation for the SSEE Eqn.~(\ref{ssee.eq}) reduces to 
\eq{\sum_{\bpp,\bpp'}\Big(A_{\bpp\bpp'}\left<\bu_{\bpp'},\chi_\br\right>_{\cO} +
  B_{\bpp\bpp'}\left<\bu_{\bpp'}^*,\chi_\br\right>_{\cO}\Big) \bu_{\bpp}(x) + \Big(C_{\bpp\bpp'}\left<\bu_{\bpp'},\chi_\br\right>_{\cO} +
  D_{\bpp\bpp'}\left<\bu_{\bpp'}^*,\chi_\br\right>_{\cO}\Big) \bu^*_{\bpp}(x) \nonumber \\
  = \mu_\br \sum_{\bpp} \Bigl( \left<\bu_\bpp,\chi_\br\right>_{\cO} \bu_{\bpp}(x)  -
  \left<\bu^*_\bpp,\chi_\br\right>_{\cO} \bu^*_{\bpp}(x) \Bigr),  \label{redssee.eq}
}where $\left<.,.\right>_\cO$ denotes the $\cL^2$ inner product in $\cO$ 
\eq{
\left<\phi_1,\phi_2\right>_\cO = \int_\cO dV_\bx\,\phi_1^*(\bx)\phi_2(\bx). \label{l2.eq}
}

We consider two cases

\noindent {{\bf (i) Finite $\mathcal L^2$ inner product:}} 
When the $\mathcal L^2$ inner product is finite  the linear independence of the $\{
\bu_\bpp\}$  gives us the coupled equations
\eq{
\sum_{\bpp'}\Big(A_{\bpp\bpp'}\left<\bu_{\bpp'},\chi_\br\right>_{\cO} + B_{\bpp\bpp'}\left<\bu_{\bpp'}^*,\chi_\br\right>_{\cO}\Big) &= \mu_\br \left<\bu_\bpp,\chi_\br\right>_{\cO},\nonumber\\
\sum_{\bpp'}\Big(C_{\bpp\bpp'}\left<\bu_{\bpp'},\chi_\br\right>_{\cO}+ D_{\bpp\bpp'}\left<\bu_{\bpp'}^*,\chi_\br\right>_{\cO}  \Big) &= -\mu_\br \left<\bu_\bpp^*,\chi_\br\right>_{\cO}.\label{redgev.eq}
}
This is  the case  when $\cO$ is compact.  

Next, assume that the $\{\bu_\bpp \}$  are  $\mathcal L^2$ orthogonal. Then 
\eq{\chi_\bap (\bx) = R \bu_\bap(\bx) + S \bu_\bap^*(\bx), \label{efun.eq}}
are eigenfunctions of Eqn.~\eqref{ssee.eq} if 
\eq{R A_{\bpp\bap} + S B_{\bpp\bap}  = \mu_{\bap} R \delta_{\bpp\bap} , \nonumber \\
   R C_{\bpp\bap} +  S D_{\bpp\bap} =  -\mu_{\bap} S \delta_{\bpp\bap}.  
   \label{musc.eq}}
 This has non-trivial solutions iff
 \eq{  (A_{\bpp\bap}- \mu_{\bap}\delta_{\bpp\bap} ) (D_{\bpp\bap}  +\mu_{\bap} \delta_{\bpp\bap} ) -
   B_{\bpp\bap}C_{\bpp\bap} =0. \label{RS.eq}
 }
 For $\bpp\neq \bap$ properties of Bogoliubov transformations implies that\footnote{This additional condition is {\it not} satisfied for example for a  causal diamond in the $d=2$ cylinder
 spacetime \cite{Mathur:2021zzl}.}\eq{|D_{\bpp\bap}|^2=|C_{\bpp\bap}|^2,\;
 \bpp\neq \bap.}
For $\bpp= \bap$, letting $A_{\bap
   \bap}=a_\bap, B_{\bap \bap}=b_\bap, C_{\bap \bap}=c_\bap, D_{\bap \bap}=d_\bap$, we see that $a_\bap, d_\bap$ are real from
 Eqn.~(\ref{abcd.eq}), so that 
 \eq{\mu_\bap^\pm=\frac{1}{2} \Biggl(1 \pm \sqrt{(1 + 2d_\bap)^2 -
     4|c_\bap|^2)}\Biggr), \label{mu.eq}}
 which is real only if \eq{(1 + 2d_\bap)^2 \geq  
   4|c_\bap|^2. \label{reality.eq}}
 This can be shown to be true using the following identity  
 \eq{\sum_{\bk} |\alpha_{\bk\bpp} - e^{i\theta} \beta_{\bk\bpp}|^2 & \geq 0 \nonumber \\
   \Rightarrow 1+2d_{\bpp} - 2 |c_{\bpp}| \cos (\theta + \theta') &\geq 0,}
 where $c_p=|c_p|e^{i \theta'}$. Taking $\theta=-\theta'$ gives us the desired relation.   
The two eigenvalues $\mu^+_\bap, \mu_\bap^-$ moreover satisfy  the relation 
 \eq{\mu^-_\bap=1-\mu^+_\bap, \label{pairs.eq}}  and therefore come in pairs $(\mu^+_\bpp,1-\mu^+_\bpp)$, as expected \cite{ssee}. 

Thus the mode-wise SSEE is 
{\eq{
\cS_\bap= \mu^+_\bap\log(|\mu^+_\bap|)+(1-\mu^+_\bap)\log(|1-\mu^+_\bap|).  \label{sp.eq}
}}

\noindent {{\bf (ii) Static Case, with compact spatial slices:}} 
Alternatively, if $\cO$ is static but with compact spatial slices but a non-compact time direction, then  $\{
\bu_\bpp(x) \}$ takes the general form 
\eq{\bu_{p\vec q}(t,\vec x) = N_{p\vec q} Z_{p\vec q}(\vec x) e^{-ipt} , \quad p > 0, \label{ssph.eq}}
where $t\in(-\infty,\infty), \, r>0$ and $N_{p\vec q}$ is a  normalisation constant with 
$p$ a continuous variable. Thus one has integrals 
over $p$  as well as summations over $l$ and $m$ in  Eqn.~(\ref{redgev.eq}).  Thus 
\begin{eqnarray} 
\left<\bu_{p\vec q},\bu_{p'\vec q'}\right>_\cO & =&  2\pi N_{p\vec q}^*N_{p \vec q'} \left < Z_{p\vec q}, Z_{p\vec q'}\right >_{\vec x}\delta(p-p') \nn \\
  \left<\bu_{p\vec q},\bu^*_{p'\vec q'}\right>_\cO & =&  0 %2\pi (N^*_{p\vec q})^2| \left < Z_{p\vec q}, Z^*_{p'\vec q'}\right >_{\vec x}\delta(p-p') \nn \\
\end{eqnarray}
where $\left<., . \right>_{\vec x}$ denotes the (finite) spatial $\cL^2$ inner product and the latter orthogonality comes from the fact that $p>0$.  
Expanding the generalised eigenfunctions $\chi_\br$ in terms of the $\{ \bu_{p,\vec q}\}$  we see that 
\begin{eqnarray} 
\left<\bu_{p\vec q}, \chi_\br\right>_\cO &=& \sum_{\vec  q'} \int dp'  \biggl( \baa_{\br p'\vec q'} \left<\bu_{ p\vec q}, \bu_{p'\vec
                                             q'}\right>_\cO \biggr) 
                                             = \sum_{\vec  q'} 2\pi \baa_{\br p\vec q'} N_{p\vec q}^*N_{p \vec q'} \left < Z_{p\vec q}, Z_{p\vec q'}\right >_{\vec x}  \nn \\
  \left<\bu_{p\vec q}^*, \chi_\br\right>_\cO &=& \sum_{\vec  q'} \int dp'  \biggl( \bbb_{\br p'\vec q'} \left<\bu_{ p\vec q}^*, \bu^*_{p'\vec
                                             q'}\right>_\cO \biggr) 
                                             = \sum_{\vec  q'} 2\pi \bbb_{\br p\vec q'} N_{p\vec q}N_{p \vec q'}^*\left < Z^*_{p\vec q}, Z^*_{p\vec q'}\right >_{\vec x} 
\end{eqnarray} 
we see the RHS of Eqn. \eqref{redssee.eq} is finite, where  the $\baa_{\br p\vec q}, \bbb_{\br p\vec q}$ are the coefficients
in the expansion of $\chi_\br$. 

Next, assume that $\{\bu_\bpp \}$  are  $\mathcal L^2$ orthogonal and that $\hW\Big|_\cO$ is  moreover block diagonal in
the $\{\bu_{p\vec q}\}$ basis,  then  
\eq{ A_{p\vec q p' \vec q'}=a_{p \vec q } \delta(p-p') \delta_{\vec q \vec q'} , \quad B_{p\vec q p' \vec q'} =b_{p\vec
    q }
  \delta(p-p') \delta_{\vec q \vec q'}, \nonumber\\
C_{p\vec q p' \vec q'}=c_{p\vec q} \delta(p-p') \delta_{\vec q \vec q'}, \quad D_{p\vec q p' \vec q'}=d_{p\vec q}
\delta(p-p') \delta_{\vec q \vec q'}.  \label{diag.eq} }
This leads to a vast simplification of    Eqn.~(\ref{redssee.eq}) which reduces to the uncoupled equations
\begin{eqnarray} 
a_{p \vec q }\lt{\bu_{p\vec q}, \chi_\br} + b_{p \vec q }\lt{\bu^*_{p\vec q}, \chi_\br} &=&  \mu_r \lt{\bu_{p\vec q},
                                                                                            \chi_\br}  \nn \\
 c_{p \vec q }\lt{\bu_{p\vec q}, \chi_\br} + d_{p \vec q }\lt{\bu^*_{p\vec q}, \chi_\br} &=& - \mu_r \lt{\bu^*_{p\vec
                                                                                             q}, 
                                                                                             \chi_\br}. 
\end{eqnarray} 
Again, the ansatz 
\eq{\chi_{p\vec q}(t,\vec x) = R \bu_{p\vec q}(t, \vec x) + S \bu_{p\vec q}^*(t,\vec x), \label{efun.eq}}
for the eigenfunctions requires that Eqn.~\eqref{RS.eq} is satisfied, as before. This yields the same form for $\mu_{plm}^\pm$ as Eqn.~\eqref{mu.eq} and hence the SSEE Eqn.~\eqref{sp.eq}.   

As we will see in the specific case of de Sitter and $d=2$ Schwarzschild de Sitter spacetimes, $\mu^+_\bap, \mu_\bap^- \not
\in (0,1)$ which is again consistent with the expectations of \cite{ssee}.  
{This analysis has been used for calculating the SSEE in de Sitter and de Sitter black hole horizons. In de Sitter spacetime we compute the SSEE of a massive scalar field in the Bunch-Davies vacuum state \cite{Bunch:1978yq} restricted to the static patch (region $I$ of Fig~\ref{fig:dS}).
\begin{figure}
\centering{\includegraphics[height=5cm]{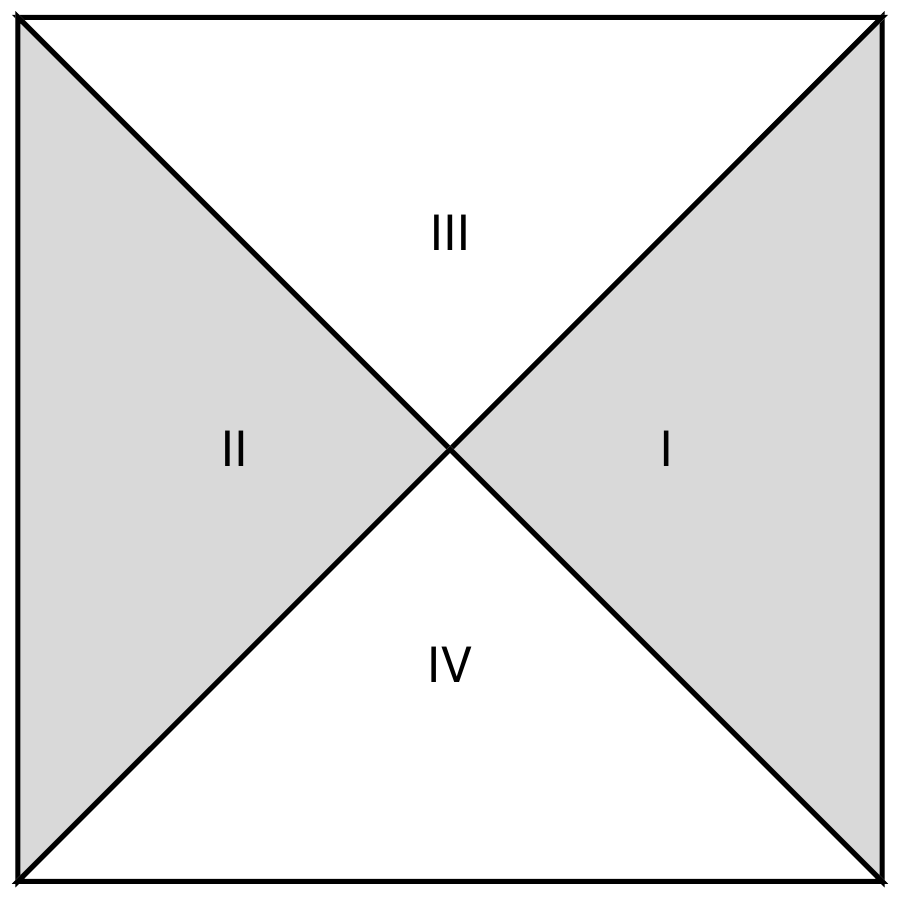}}
\caption{A Penrose diagram for de Sitter spacetime. }
\label{fig:dS}
\end{figure}
The modes we use in the static patch are normal modes found by Higuchi \cite{Higuchi:1986ww} which are also $\cL^2$ orthogonal. Following the calculations of \cite{Higuchi:2018tuk} we find in \cite{ourwork2} that
\eq{A_{pp'}=\frac{\delta(p-p')}{1-e^{-2\pi p}},\;\;D_{pp'}=\frac{\delta(p-p')}{e^{2\pi p}-1}\;\;\text{and}\;\;B_{pp'}=C_{pp'}=0.}
which then leads to the mode-wise contribution to SSEE
\eq{\cS_{plm} = -\log(1-e^{-2\pi p}) - \frac{e^{-2\pi p}}{1-e^{-2\pi p}}\log e^{-2\pi p},}
which agrees with the von Neumann entropy evaluated by \cite{Higuchi:2018tuk}. In particular, the result is independent
of the mass. 

In Schwarzschild de Sitter spacetime we compute the SSEE of a massless scalar field restricted to one of the static patches (region $I$ of Fig.~\ref{fig:sds})\cite{ourwork2}. The field is assumed to be in the Kruskal vacuum state which is defined across the black hole horizon. We use static modes to expand the restriction of the Kruskal Wightman  function to the static patch.
\begin{figure}
\centering{\includegraphics[]{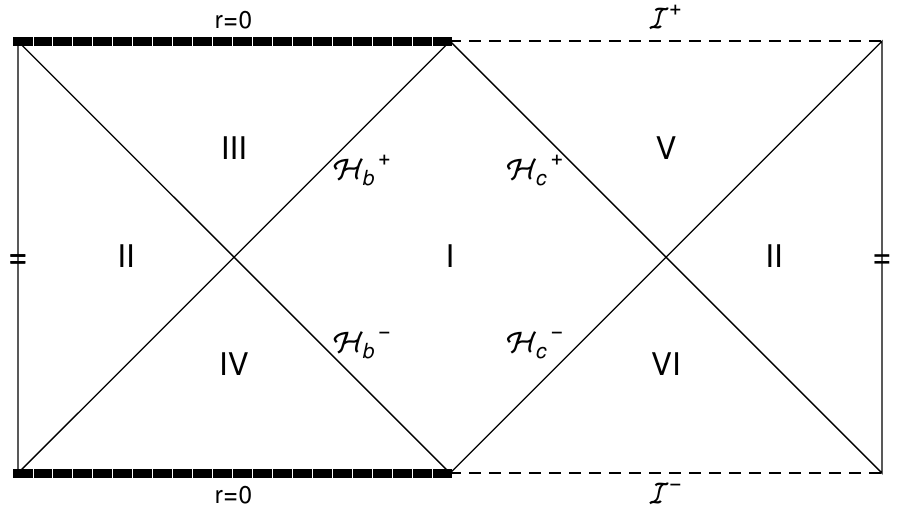}}
\caption{The Penrose diagram for the $d>2$ Schwarzschild de Sitter spacetime, where each point represents an
    $\ms^{d-2}$ and each horizontal slice represents an $\ms^{d-2}\times\ms^1$. Region $I$ and $II$ are the static patches,
    and $\cH_b^{\pm}$ and $\cH_c^{\pm}$ are the black hole and the cosmological horizons respectively.}
\label{fig:sds}
\end{figure}
The massless scalar field modes (Kruskal and static) are not known in full static patch but since the SSEE depends only on the Bogoliubov transformation of these modes, the knowledge of the modes in a neighbourhood of a Cauchy hypersurface is sufficient to calulate the SSEE. We use the past boundary conditions for the static and the Kruskal modes, since this defines the Klein Gordon norm on the 
limiting initial null surface $\cH_b^- \cup \cH_c^-$ in Region I. As shown in \cite{ourwork2} we find that
\eq{A_{pp'}=\frac{\delta(p-p')}{1-e^{-2\pi p/\kappa_b}},\quad D_{pp'}=\frac{\delta(p-p')}{e^{2\pi p/\kappa_b}-1},\;\;\text{and}\quad B_{pp'}=C_{pp'}=0.}
which leads to the mode-wise contribution to SSEE
\eq{\cS_{plm} = -\log(1-e^{-2\pi p/\kappa_b}) - \frac{e^{-2\pi p/\kappa_b}}{1-e^{-2\pi p/\kappa_b}}\log e^{-2\pi p/\kappa_b}.}
}
where $\kappa_b$ is the surface gravity of the black hole horizon.

\begin{figure}[h!]
	\centering
	\includegraphics[scale=0.8]{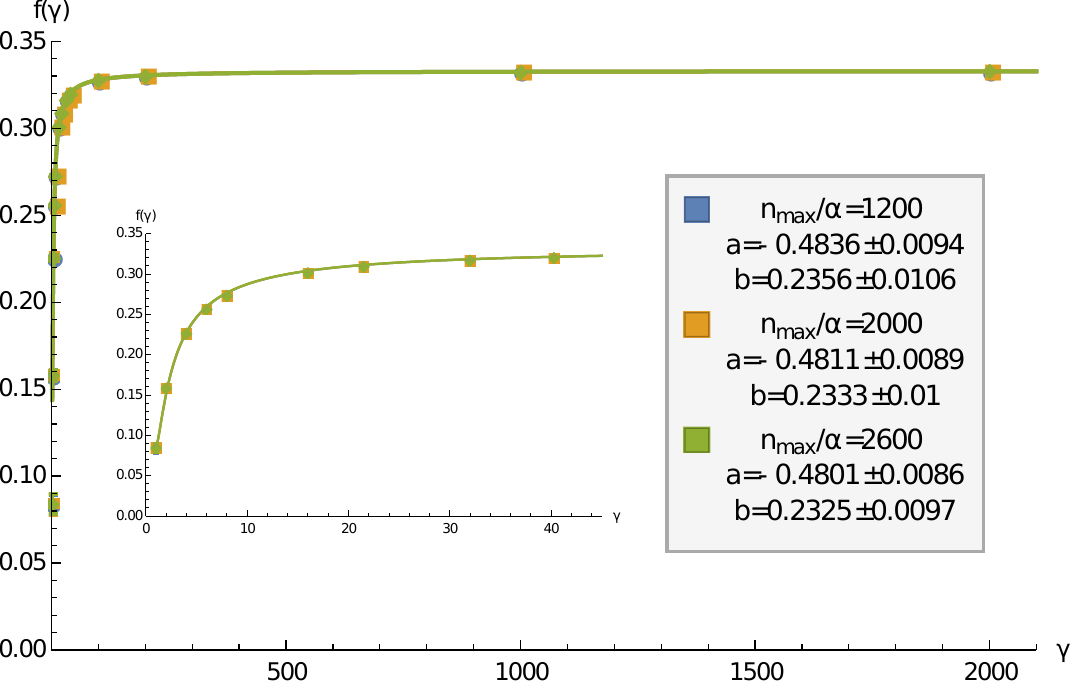}
	\caption{A plot of $f(\gamma)$ vs. $\gamma$ for different values of $n_\mx/\alpha$, fitted to 
		$0.33+a/\gamma+b/\gamma^2$.  
		The inset figure shows the smaller $\gamma$ values.} \label{fgamma.fig} 
\end{figure}

{We now move on to the next example which is the SSEE for a massless scalar field in a causal diamond inside a slab of
  2d cylinder spacetime
  \begin{equation}
ds^2= -dt^2+d\phi^2,\quad \phi\sim\phi+2\pi. 
    \end{equation} 
We begin with the Fewster-Verch SJ vacuum $\wsj$ in a slab of $t \in [-T,T]$ and restrict it to a causal diamond
$\diam_2$ as shown in Fig~\ref{cyl.fig}.
\begin{figure}[htb]%{0.5\textwidth}
	\centering{\includegraphics[height=6.6cm]{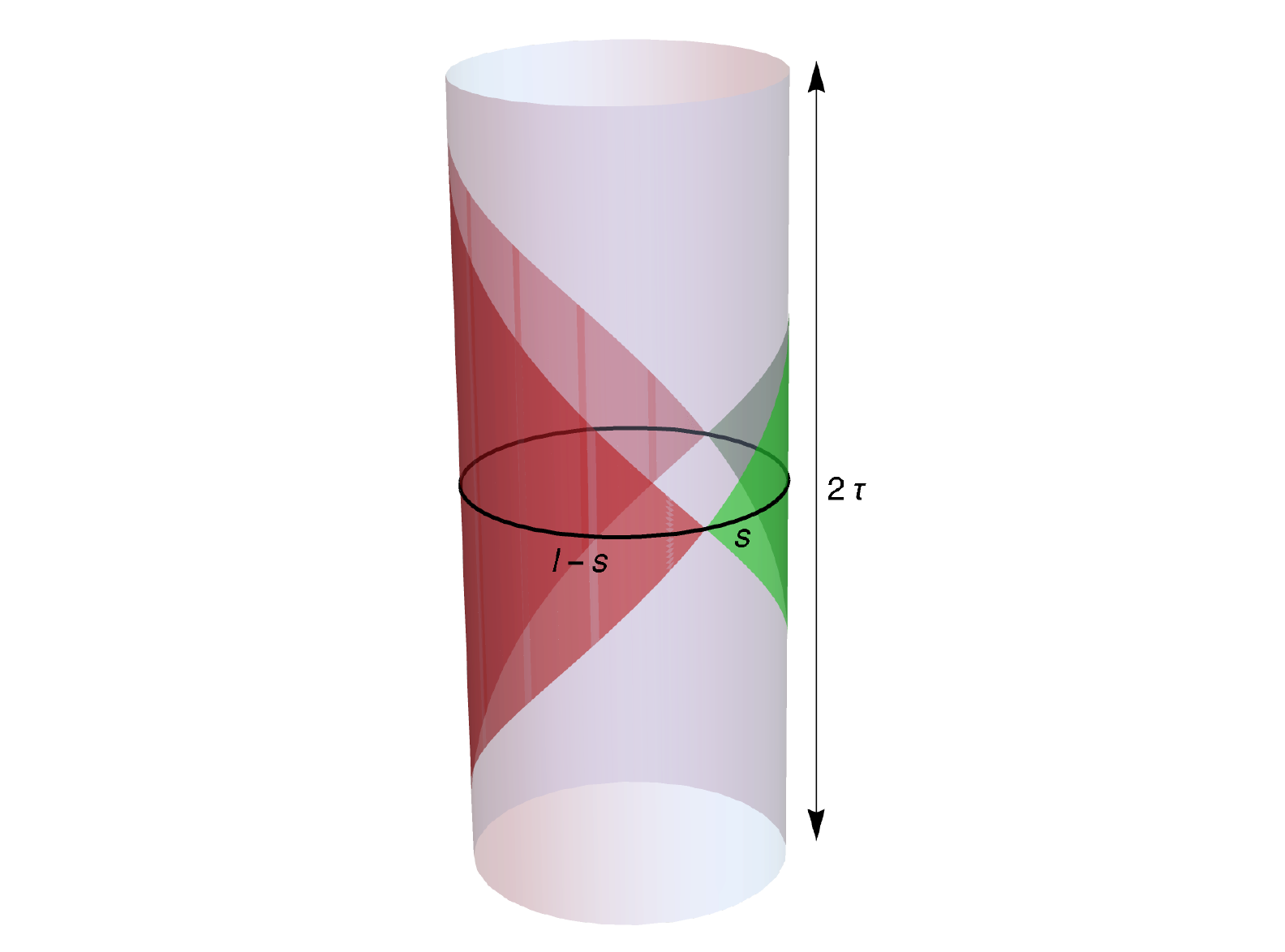}}
	\caption{The causal diamond and its causal complement in the $d=2$ cylinder spacetime.}
	\label{cyl.fig} 
\end{figure}
Let $L$ denote the circumference of the cylinder and $\ell$ the proper time of the causal diamond.    Unlike the cases we have just studied,  the restricted SJ Wightman function is not block diagonal with
respect to  the SJ modes in the diamond.  Hence the analysis discussed above cannot be extended to get the generalised
eigenvalues of the form given by Eqn.~\eqref{mu.eq}.  Instead, as shown in \cite{Mathur:2021zzl} we use  a
combination of analytical and numerical techniques to calculate the generalised spectrum.

$\hW\big|_{\diam_2}$ can be expanded in the SJ basis or eigenbasis of  $i\hD$ in diamond, but in general this
expression does not have a closed form. Solving the generalised eigenvalue equation numerically therefore requires an additional
cut-off.  In \cite{Mathur:2021zzl} it was noticed that when the $\gamma=2 T/L $ is a half
integer and the ratio $\alpha =\ell/L$  is rational then $\hW\big|_{\diam_2}$  simplifies considerably into a closed
form expression. We refer the reader to \cite{Mathur:2021zzl} for details.  In this case, the SSEE can be calculated 
numerically by imposing a {\it covariant}  UV cut-off in the SJ spectrum in $\diam_2$ which renders the problem finite.

We plot the SSEE for different values of the parameters $\alpha,\gamma$ and the cut-off $n_\mx$ %as shown in Fig.~\ref{fig:cyleevsalpha} and \ref{fig:cyleevskmax}.
\begin{comment}
\begin{figure}
\centering{\includegraphics[height=4cm]{EE_vs_alpha_nmaxbyalpha2600_for_suppliment.pdf}}
\caption{$\ssee$ vs $\alpha$ for different $\gamma$ with $n_\mx/\alpha=2600$ fitted to $\ssee=a\log(\sin(\pi\alpha))+b$. The fit parameters are shown in the table.}
\label{fig:cyleevsalpha}
\end{figure}
\begin{figure}
\centering{\includegraphics[height=4cm]{EE_vs_kmax_gamma200_for_suppliment.pdf}}
\caption{A log-linear plot of  $\ssee$ vs $n_\mx/\alpha$ for different $\alpha$ with $\gamma=200$ fitted to $\ssee=a\log\,(n_\mx/\alpha)+b$. The fit parameters are shown in the table which show $a \sim 1/3$. This is also true for other values of $\gamma$. The curves for  complementary values of $\alpha$ are indistinguishable.}
\label{fig:cyleevskmax}
\end{figure}
\end{comment}
and use the best fit to obtain SSEE, which is of the form
%The SSEE thus obtained is of the form
\eq{\sseecyl = \frac{c(\gamma)}{3}\log\left(\frac{l}{\pi\epsilon}\right) + f(\gamma)\log\left(\sin(\pi \alpha)\right) + c_1(\gamma),\label{sseecyl.eq}}
where $\epsilon$ is the UV cut-off, which in terms of the cut-off in SJ spectrum is given by
\eq{\epsilon = \frac{l\alpha}{2\sqrt{2}\pi n_\mx} = \frac{l(1-\alpha)}{2\sqrt{2}\pi n_\mx'},}
where $n_\mx$ is the cut-off in the SJ spectrum\footnote{One may refer to \cite{Mathur:2021zzl} for details.}. It is
shown numerically in \cite{Mathur:2021zzl} using best fit curves that $c(\gamma)=1$, $f(\gamma)$ converges to unity for
large enough $\gamma$ as shown in Fig.~\ref{fgamma.fig}, and therefore the SSEE in Eqn.~\eqref{sseecyl.eq} reduces to
the exact Calabrese-Cardy form in this limit. It is also clear from Eqn.~\eqref{sseecyl.eq} that the SSEE for a complimentary pair of diamonds are equal. $c_1(\gamma)$, which is a non-universal term in the Calabrese-Cardy entropy formula, inreases with $\gamma$ logarithmically.}

Despite its asymptotic behaviour it is interesting that the coefficient $f(\gamma)$ is  
not ``universal''.   Importantly the SJ vacuum in the slab itself changes with $\gamma$ and is a pure state in the
infinite cylinder,  though not its ground state. Thus, one can view the SSEE that we have calculated to be that for a 
family of pure states in the infinite cylinder rather than that of its vacuum.  

\section{SSEE in the Causal Set} 
\label{causalset}

Associated with  every causal spacetime $(M,g)$,  is  a classical  ensemble of  causal sets $\{ C\}$, where each $C$ is
obtained via a Poisson sprinkling at density $\rho$ into $(M,g)$, with the order relations given by the continuum
causality relations.  For such a random discretisation, the   probability of finding $n$ elements in a spacetime region
$V$ is  
\begin{equation}
	P_V(n) = \frac{(\rho V)^n}{n!} e^{-\rho V} 
\end{equation} 
and we have a mean number to volume correspondence
\begin{equation}
	\av{N} = \rho V.
\end{equation}

\begin{figure}
	\centering
	\includegraphics[width=0.75\textwidth]{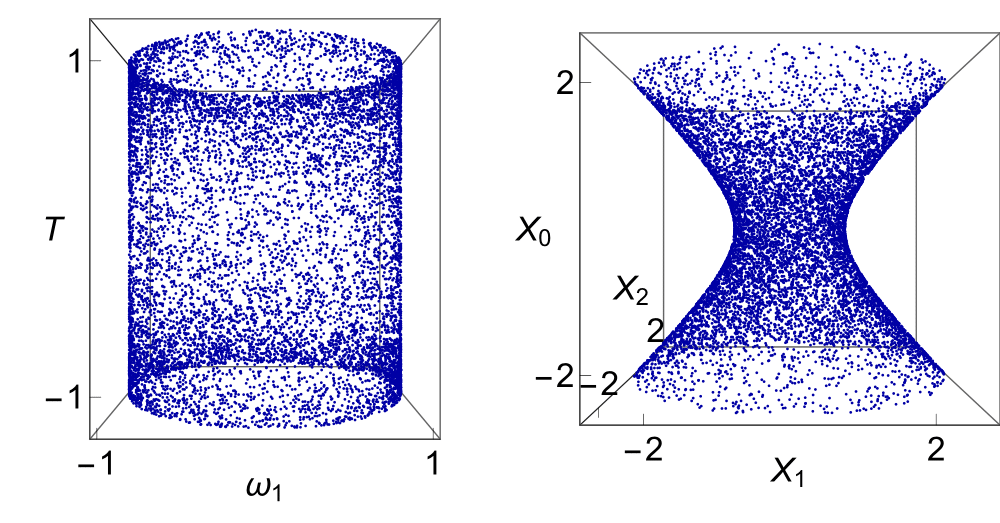}
	\caption{A causal set approximated by de Sitter spacetime. The representation on the left is in conformal coordinates and on the right is in usual hyperbolic coordinates.}
	\label{dscauset}
\end{figure}
Fig \ref{dscauset} is an example of a causal set that is approximated by de Sitter spacetime. Because of this discrete Poisson
randomness, the causal set discretisation is covariant and locally Lorentz invariant, thus making it an ideal candidate
for regulating the infinities of quantum field theory.

As before we concern ourselves only with the free scalar field theory, and construct the quantum vacuum via the
Sorkin-Johnston procedure, which requires us to first obtain the advanced and retarded Green's functions on $C$. We will
need to define first the causal matrix
\begin{equation}
	C_0(x,x') \equiv
	\left\{
	\begin{array}{ll}
		1  & \mbox{if } x' \prec x \\
		0 & \mbox{} \mathrm{otherwise}
	\end{array}
	\right.
\end{equation}
and the link matrix
\begin{equation}
	L_0(x,x') \equiv 
	\left\{
	\begin{array}{ll}
		1  & \mbox{if } x' \prec x\,\,\mathrm{and} \,\,|[x,x']|=0 \\
		0 & \mbox{} \mathrm{otherwise}
	\end{array}
	\right. 
\end{equation}
where  $[x,x'] \equiv \{ z\in C \,|\, x'\prec z \prec x \}$.  The $k$-chain matrix is then  $C_k \equiv  C_0^k$ and the
$k$-link matrix is then  $L_k \equiv  L_0^k$. %\mnote{check if its k or k-1. N: k is correct.} 
As shown in \cite{johnston}, by comparing with the continuum in $\mink^2$ and $\mink^4$, the massless causal set Green's functions
can be written  in  terms of these matrices 
\begin{equation}
	\label{massless2d} 
	K^{(2)}_0(x,x')\equiv  \frac{1}{2}C_0(x,x'), \quad K^{(4)}_0(x,x')= \frac{1}{2 \pi} \sqrt{\frac{\rho}{6}} L_0(x,x'), 
\end{equation} 
respectively, and their massive counterparts by
\begin{equation}
	\label{hopstop}
	K^{(2)}_m(x,x') =  \sum\limits_{k=0}^\infty  \biggl(-\frac{m^2}{\rho}\biggr)^k\biggl(\frac{1}{2}\biggr)^{k+1} C_k(x,x'), 
\end{equation}
and 
\begin{equation}
	\label{discreteGF4d}
	K^{(4)}_m(x,x') \equiv   \sum\limits_{k=0}^\infty
	\biggl(-\frac{m^2}{\rho}\biggr)^{k}\biggl(\frac{1}{2\pi}\sqrt{\frac{\rho}{6}}\biggr)^{k+1} L_k(x,x'). 
\end{equation}
respectively. In \cite{Nomaan:2017bpl} it was shown that this simple form of the causal set Green's  function is still valid in
the Riemann normal neighbourhoods of all $d=2$ spacetimes and those of $d=4$ spacetimes with  $R_{ab}\propto g_{ab}$. Of relevance to this
work is the result of \cite{Nomaan:2017bpl} that this is also the Green's function for $d=2$ and $d=4$  de Sitter and anti de
Sitter spacetimes.

Using the Sorkin-Johnston formulation described in Section~\ref{SSEE.sec}, one can then construct  the scalar  quantum field  vacuum on causal sets approximated by this above
set of  of spacetimes. This is the starting point for finding the SSEE  on causal sets.

As in the continuum, we are interested in finding the quantum scalar field SSEE in a causally convex subcausal set  $C'
\subset C$ with respect to its causal complement.  Starting with the SJ vacuum $\hW$ in $C$ which is a pure state, we
wish to find the SSEE for the mixed state $\hW_{\cC_\cO}$ obtained by simply restricting  to the region $C'$.

\begin{figure}[!h]
	\begin{subfigure}{0.3\textwidth}
		\includegraphics[width=\linewidth]{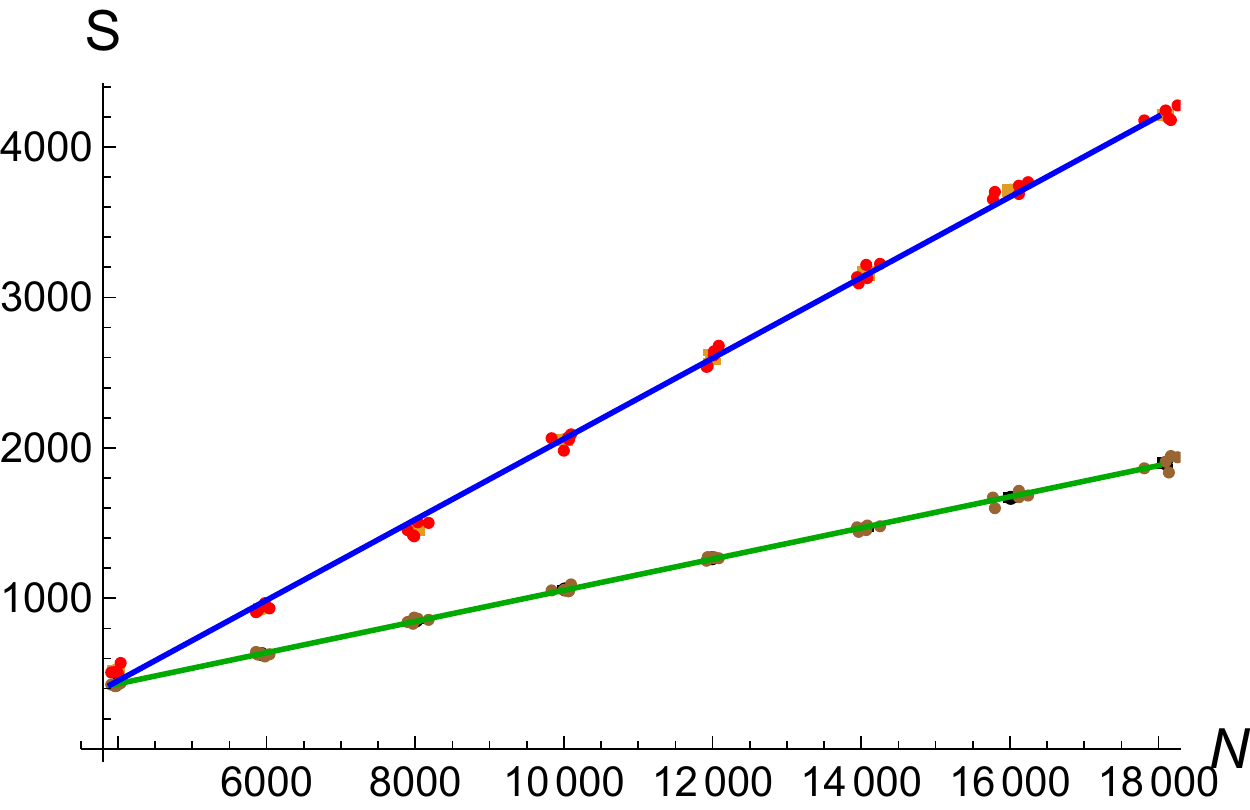}
		\caption{causal diamonds in $\mink^4$ }
	\end{subfigure}
	\begin{subfigure}{0.3\textwidth}
		\includegraphics[width=\linewidth]{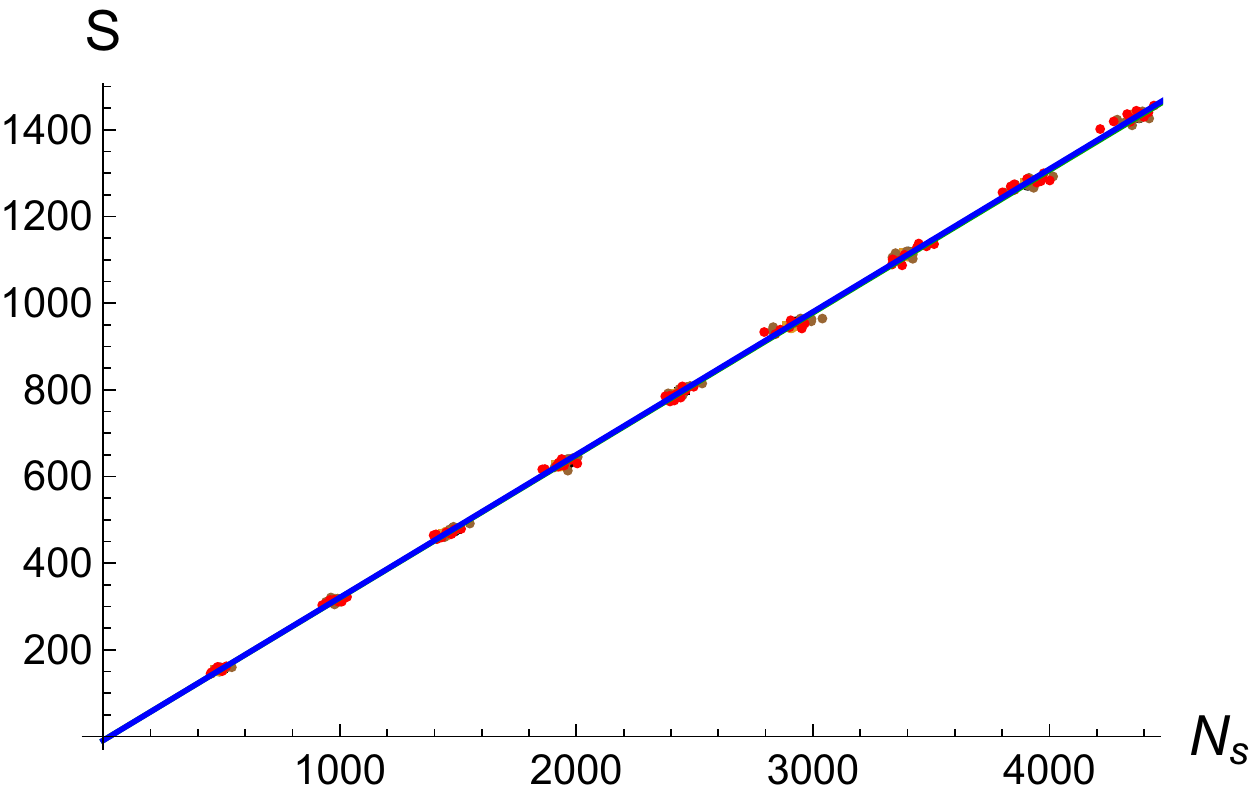}
		\caption{$d=2$ de Sitter }
	\end{subfigure}
	\begin{subfigure}{0.3\textwidth}
		\includegraphics[width=\linewidth]{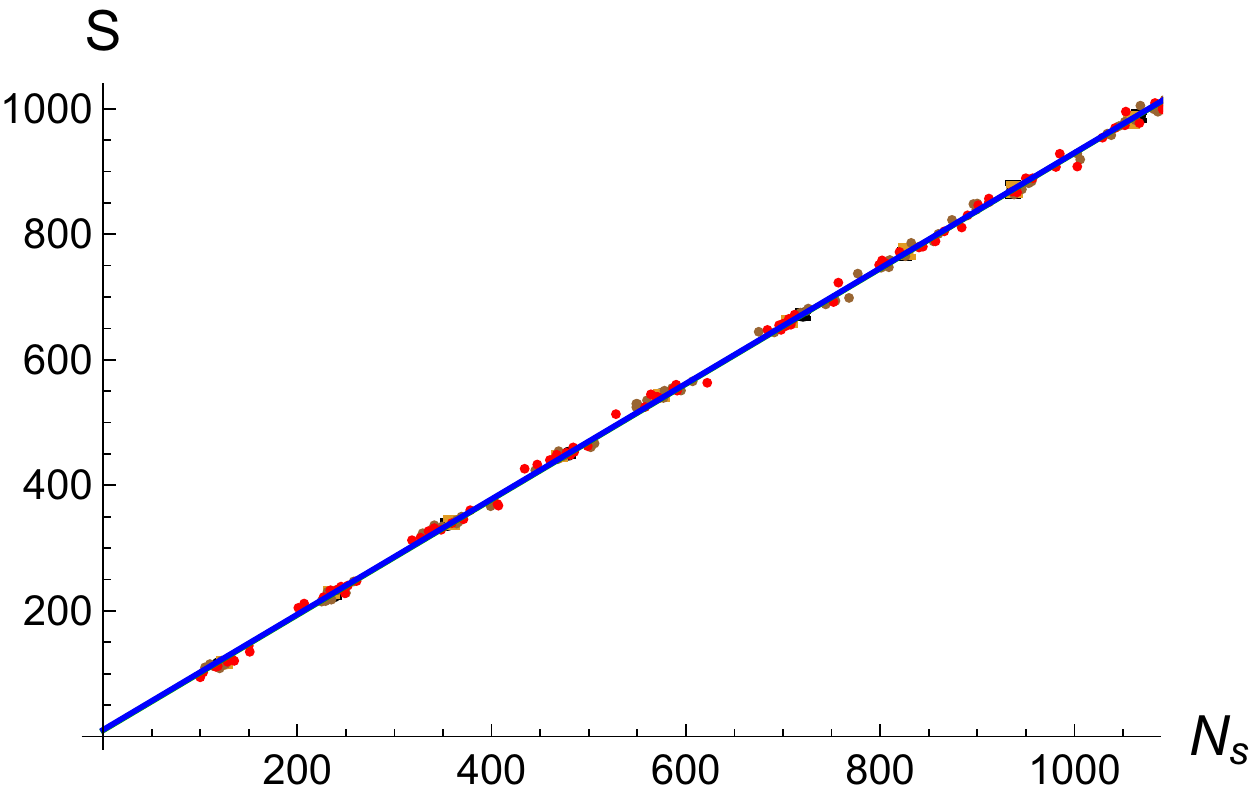}
		\caption{$d=4$ de Sitter}
	\end{subfigure}
	\caption{SSEE vs. $N$ for the horizon EE of a causal set approximated by different spacetime regions. Green and blue represent the data for the two complemetary regions. Note that the complemetarity in the $\mathbb{M}^4$ case is not obvious because, unlike the de Sitter case, the volumes and geometry of the complementary regions are not the same.}
	\label{vol_law}
\end{figure}

The causal set SSEE mimics the one in the continuum with the added simplicity that operators and functions are realized
as matrices, so that 
\begin{equation}
	\sum_{e'\in C'}W_{ee'}\Psi_{e'}  = i  \mu  \sum_{e'\in C'}\Delta_{ee'}\Psi_{e'}, \quad \Psi \nin \mathrm{ker}\,{i \Delta}, \quad
	\mathcal S= \sum_{\mu} \mu \ln |\mu|.
	\label{causet.ssee.eq} 
\end{equation} 
Because of the finiteness of the matrix it is already obvious that the SSEE is finite.  In Fig \ref{vol_law} we show the
behaviour of the SSEE for different manifold-like causal sets. What is obvious from all of these is that rather than an
area law, the SSEE explicitly follows a volume law \cite{soryaz,syx-entropy}.

\begin{figure}[!h]
\centering
 	\begin{subfigure}{0.3\textwidth}
 	\includegraphics[width=\textwidth]{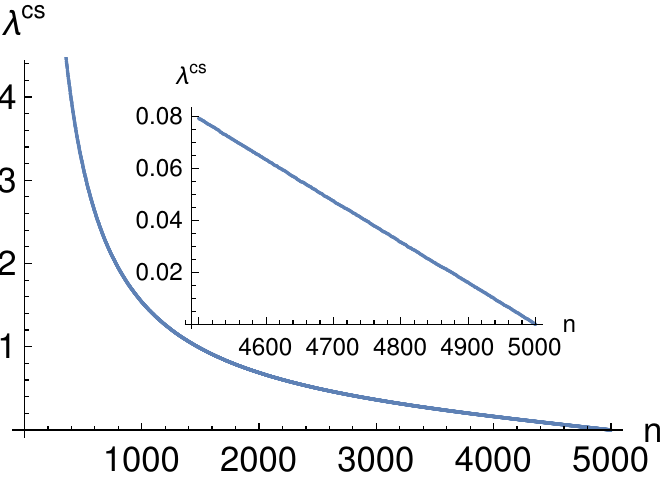}
 	\caption{Causal diamond in $\mink^2$}
 	\end{subfigure}
 	\begin{subfigure}{0.3\textwidth}
 	\includegraphics[width=\textwidth]{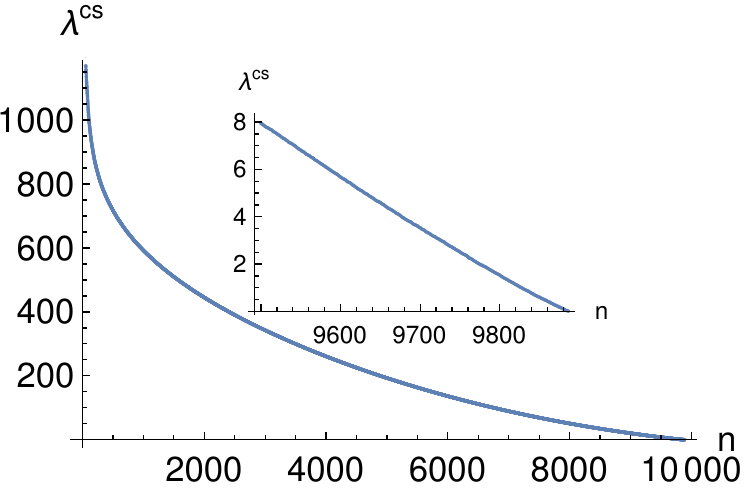}
 	\caption{Causal diamond in $\mink^4$}
 	\end{subfigure}\\
 	\begin{subfigure}{0.3\textwidth}
 	\includegraphics[width=\textwidth]{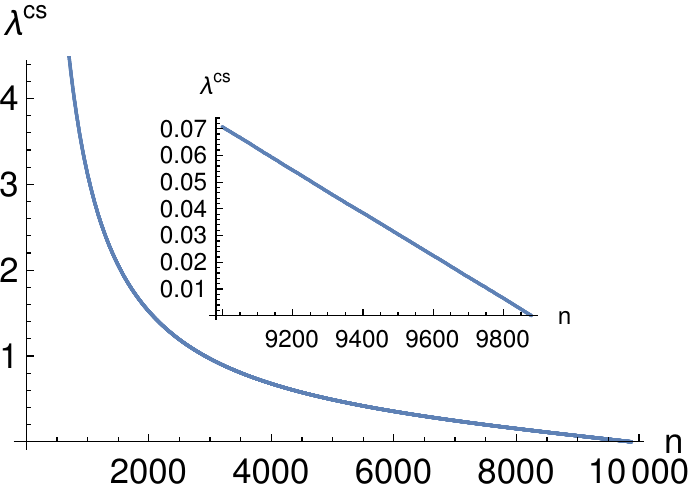}
 	\caption{$d=2$ de Sitter slab spacetime}
 	\end{subfigure}
 	\begin{subfigure}{0.3\textwidth}
 	\includegraphics[width=\textwidth]{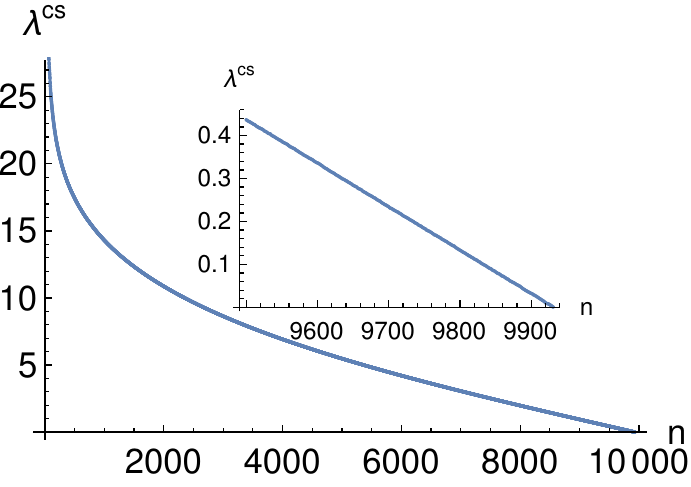}
 	\caption{$d=4$ de Sitter slab spacetime}
 	\end{subfigure}
 	\caption{The spectrum of $i\hD$ for massless scalar field. For large $n$ the spectrum varies linearly with $n$
          as shown in the inset figures. The continuum spectrum on the other hand follows a power law $\propto n^{-\alpha}$ with an
        infinitely long tail.}
 	\label{causetspectrumlargen.fig}
\end{figure}

\begin{figure}[!h]
\centering
 	\begin{subfigure}{0.3\textwidth}
 	\includegraphics[width=\textwidth]{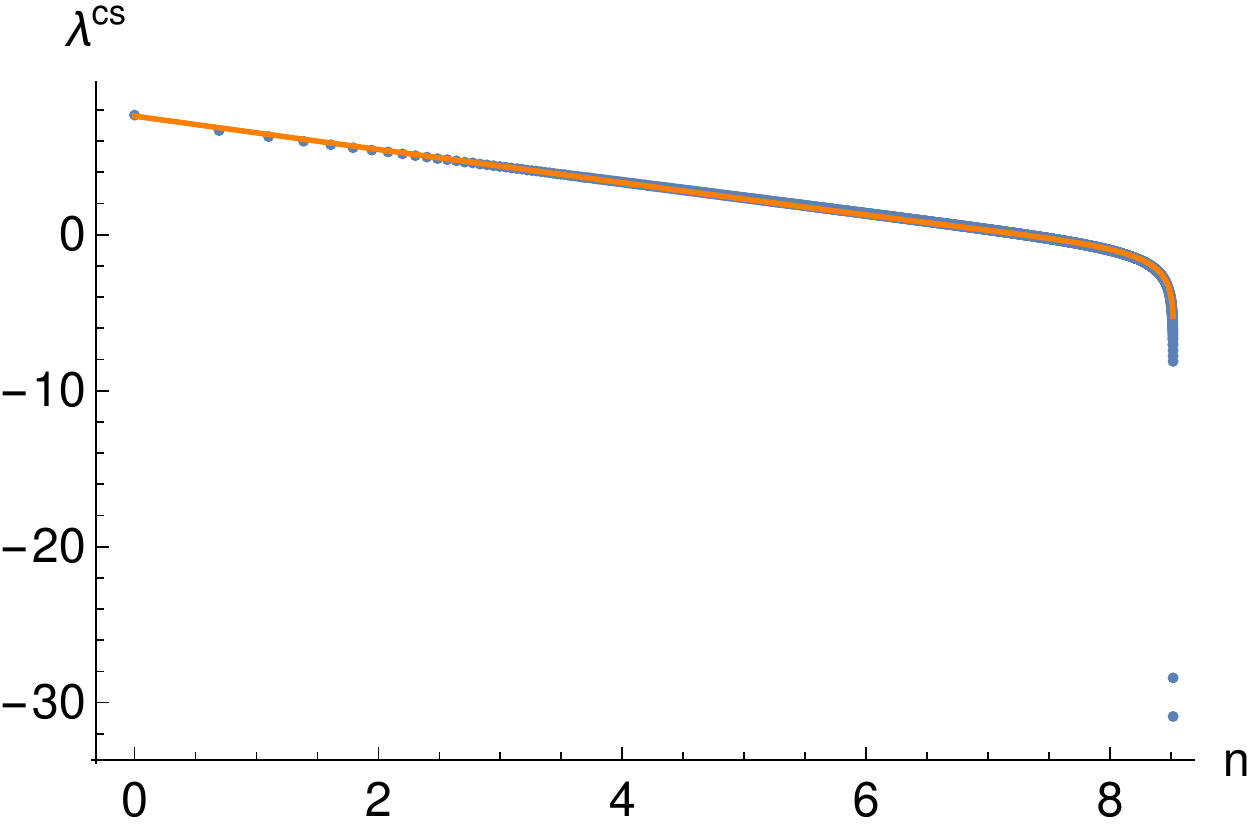}
 	\caption{Causal diamond in $\mink^2$}
 	\end{subfigure}
 	\begin{subfigure}{0.3\textwidth}
 	\includegraphics[width=\textwidth]{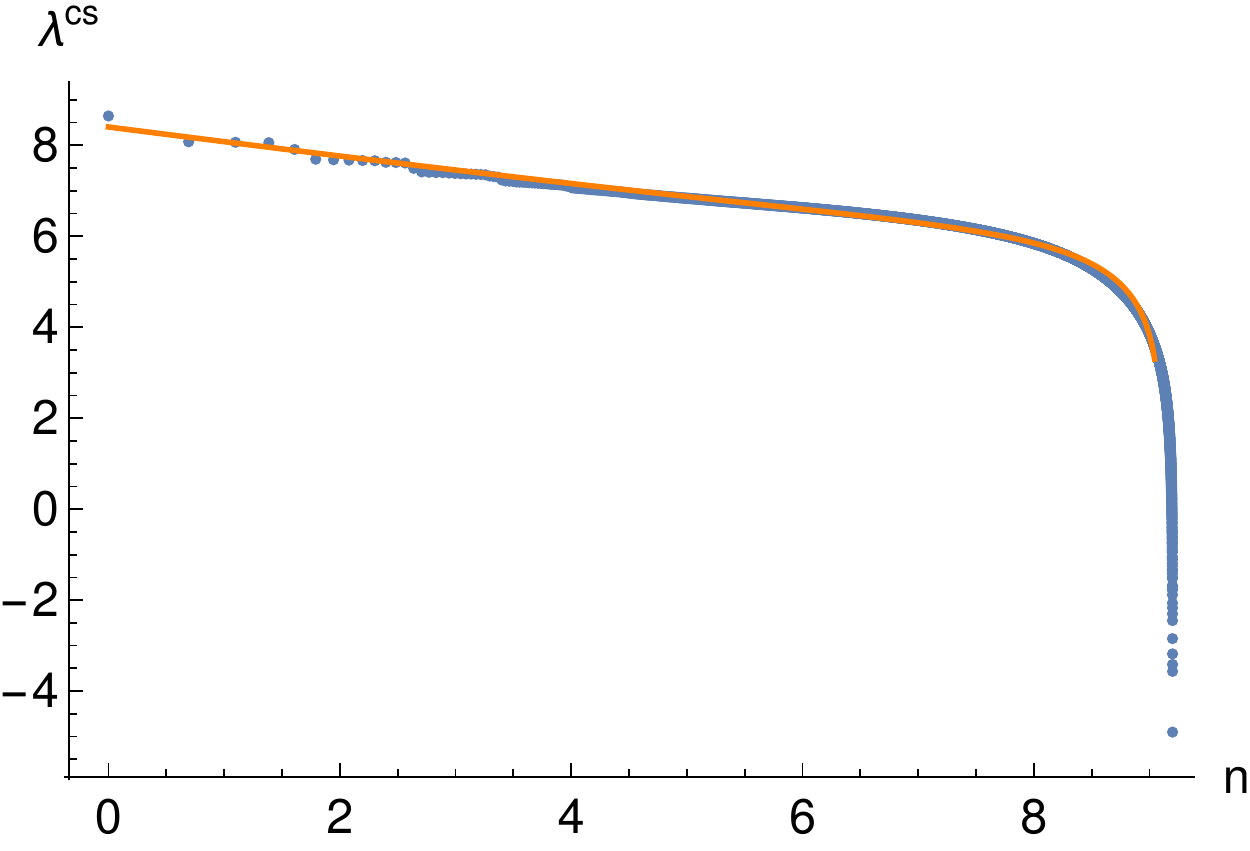}
 	\caption{Causal diamond in $\mink^4$}
 	\end{subfigure}\\
 	\begin{subfigure}{0.3\textwidth}
 	\includegraphics[width=\textwidth]{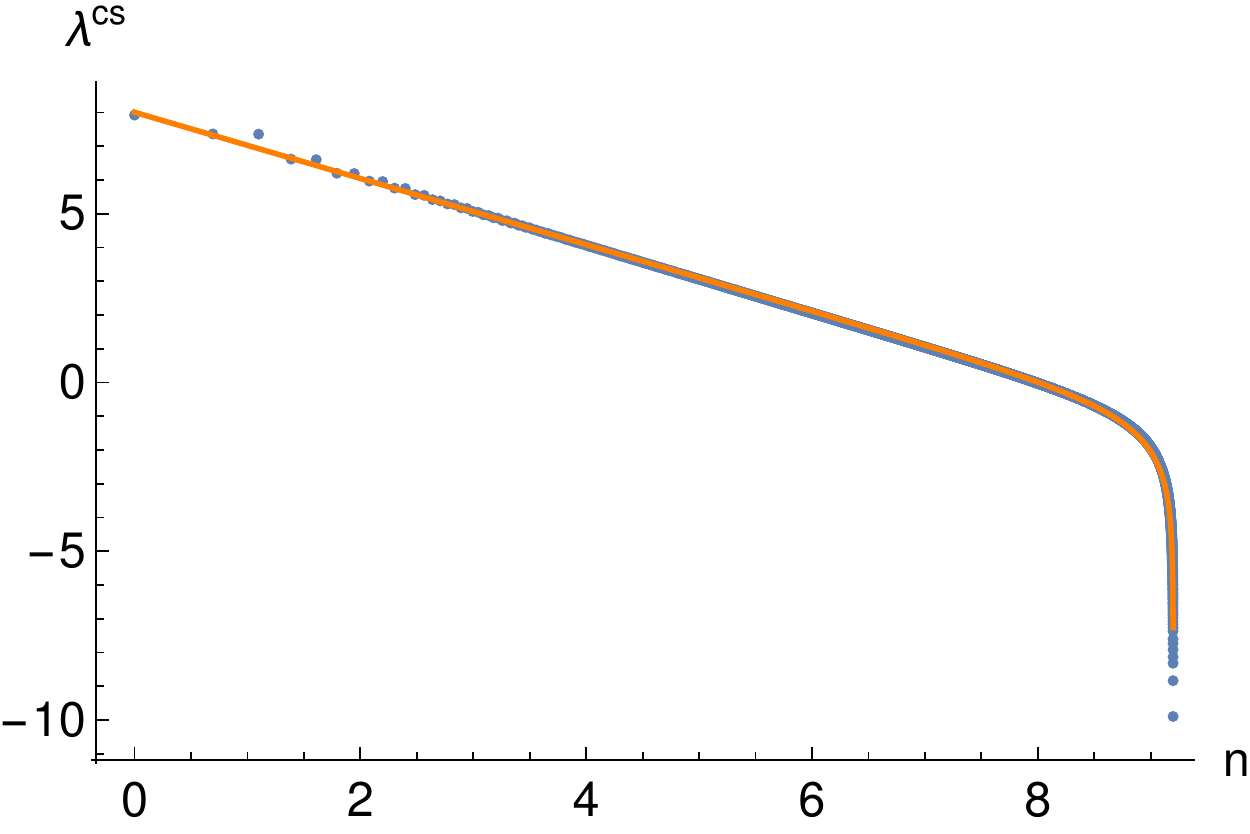}
 	\caption{$d=2$ de Sitter slab spacetime}
 	\end{subfigure}
 	\begin{subfigure}{0.3\textwidth}
 	\includegraphics[width=\textwidth]{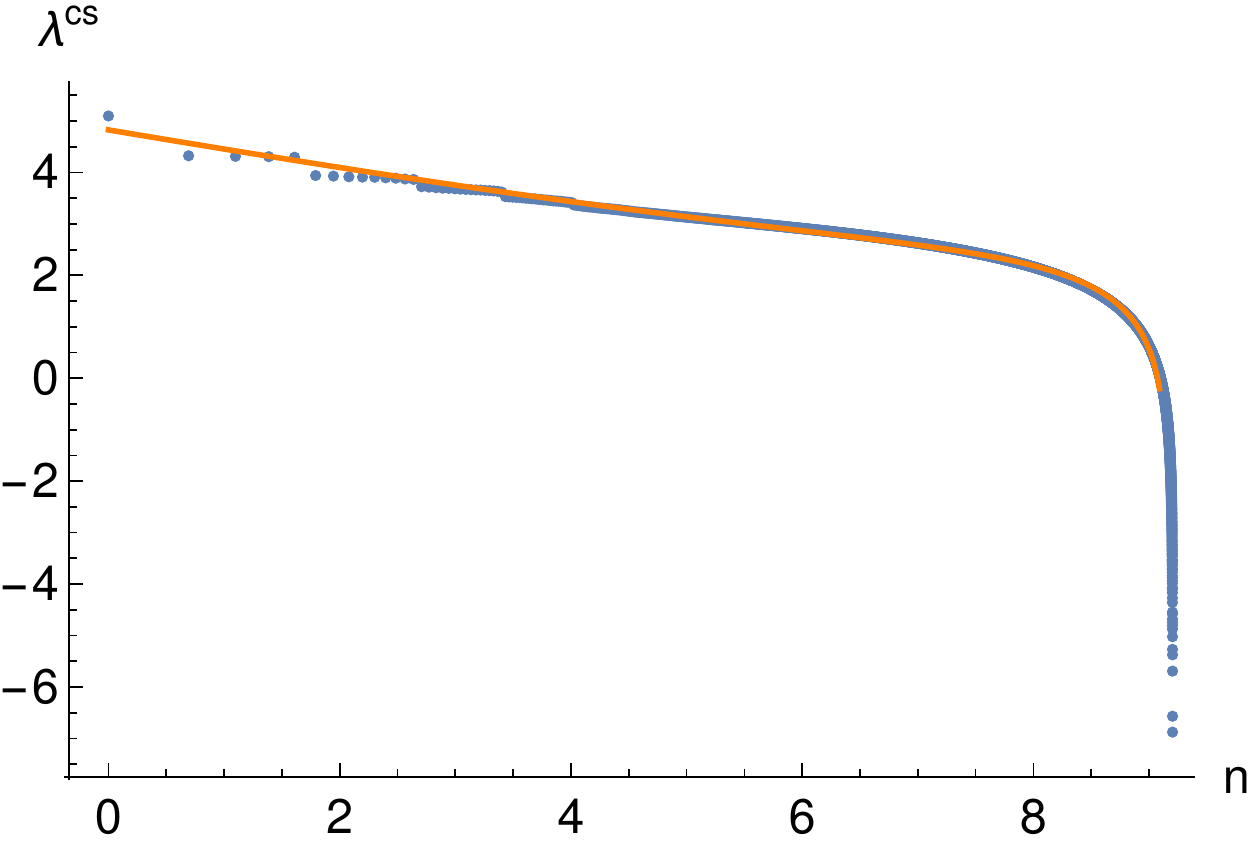}
 	\caption{$d=4$ de Sitter slab spacetime}
 	\end{subfigure}
 	\caption{A log-log plot of the spectrum of $i\hD$ for massless scalar field. It follows $\lambda^{cs} = \dfrac{\beta_1}{n^{\alpha_1}} + \alpha_2 n + \beta_2$ as shown in orange.}
 	\label{causetspectrumlog.fig}
\end{figure}

There are two ways to interpret this result : (i)  that the SSEE on causal sets is not a good measure of
entanglement and should be modified somehow by inserting an additional cut-off or truncation in the spectrum because the
quantum field theory in the deep UV cannot be trusted or (ii) that volume laws are natural for non-local field theories
and since causal sets are fundamentally non-local, it is to be expected that one should get a volume law. 
\begin{figure}[!h]
	\centering
	\begin{subfigure}{0.45\textwidth}
	\includegraphics[width=\textwidth]{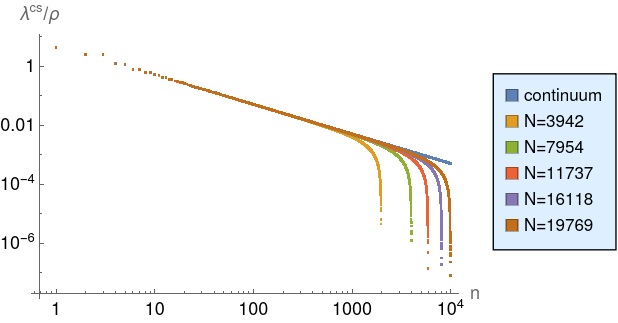}
	\caption{$d=2$ de Sitter}
	\end{subfigure}
	\begin{subfigure}{0.45\textwidth}
	\includegraphics[width=\textwidth]{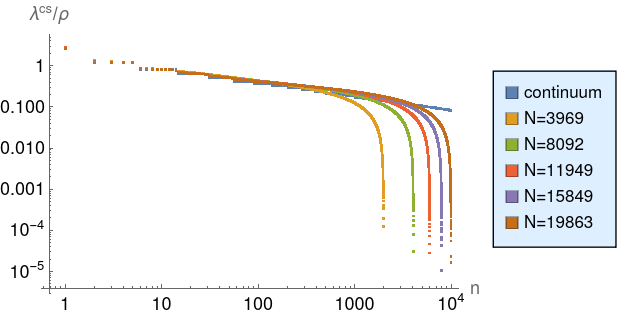}
	\caption{$d=4$ de Sitter}
	\end{subfigure}
	\caption{\small A log-log plot of the SJ spectrum wavelength $\lambda$ versus quantum number $n$. The continuum
          spectrum shown in blue exhibits a scaling behaviour while the coloured plots are the causal set spectrum for
			different discreteness scales.}
	\label{sjspectrum_2dds}
\end{figure}
In the remaining part of this section  we explore the first of these options and leave the second to the discussions 
section.  The former gives us a way out of a volume law, by mimicking the UV cut-off required in the continuum. For this, it 
is instructive to examine Fig \ref{SJspectrum.fig} where the SJ spectrum in a continuum $d=2$ causal diamond  is plotted
alongside that in the causal set at different sprinkling densities. Fig \ref{sjspectrum_2dds} shows a similar plot for de Sitter
	spacetime.  The continuum spectrum in both cases follows a scaling behaviour 
\begin{equation}
	\lambda= \frac{b}{n^{\alpha}} 
\end{equation} 
for some $\alpha$.  The corresponding causal set spectrum $\rho^{-1} \lambda^{cs}$ on the other hand trails
the scaling behaviour upto a ``knee'' beyond which its UV behaviour follows a wholly different, non-scaling
form\footnote{$\lambda^{cs}$ has the same physical  dimensions as $i\Delta$   while  $\lambda$ has the  physical dimensions of $[length]^2$.}.
Instead, a strong linear behaviour begins to dominate at  large $n$ as shown in Fig.~\ref{causetspectrumlargen.fig}, so that we may write 
\begin{equation}
  \lambda^{cs} \sim \begin{cases}\dfrac{\beta_1}{n^{\alpha_1}}, \quad n< n_0 \\ 
\- \alpha_2 n + \beta_2, \quad n>n_0.\end{cases}
\end{equation}
In the log-log plots of Fig ~\ref{causetspectrumlog.fig}, the full spectrum is shown to be roughly well modeled by a sum of these
functions. An important qualitative feature of the spectrum is that even for large $n$ or the deep UV, far being random
or chaotic, it is relatively smooth, rapidly approaching  zero, as it should in any finite
theory. 
Beyond this knee, the spectrum consists  of a large number of small but non-zero
eigenvalues which dominate the SSEE.  

In \cite{soryaz} the SSEE was recalculated in the $d=2$ nested causal set diamonds by employing a truncation of the causal
set  spectrum at the ``knee''.  Since this is at one end of the continuum-like scaling regime of the spectrum, it mimics
the UV cut-off in  the continuum SJ spectrum. However, it was shown in \cite{soryaz} that this
is not in itself enough -- one has to do a ``double truncation'' by truncating the SJ spectrum in the larger diamond and
then  truncating the SJ spectrum once again  in the smaller diamond. This removes the  additional high energy modes that
creep back in after  the first truncation.

In the causal diamond, the knee in the causal set  SJ spectrum can be found simply by comparison with the continuum
  eigenvalues $\lambda=L/k$,
where $2 \sqrt{2} L$ is the proper
time of the diamond and where  $k\sim \dfrac{n\pi}{L}$ for large  $k$.  Causal set discreteness determines a ``smallest
wavelength''  $\nu_{\text{min}}\sim\rho^{-1/2}=2L/\sqrt{N} =2\pi/k_\text{max}=2L/n_\text{max}$ or
$n_\text{max}=\sqrt{N}$.  Thus, a reasonable  cut-off is
$\lambda^\text{cs}_{\text{min}}=\rho\lambda_\text{min}=\sqrt{N}/4\pi$. In Fig \ref{SJspectrum.fig}
this value corresponds roughly to the knee of the causal set SJ spectrum. %\red{N: can we see this from Fig 1?}

In the absence of knowledge of the continuum spectrum, $n_{\text{max}}$ needs to be obtained from more general
arguments. In the continuum, 
we expect the Cauchy hypersurface to contain all information about the QFT. Even though the dimension of $i \hD$ is $N$,
we know that the space of independent solutions of the equations of motion is spanned by $\im(i \hD)$ \cite{wald} and
that this picture should be consistent with the continuum. Therefore the dimension of $\im(i \hD)$ must be related to
the spatial volume of the Cauchy hypersurface which, for the symmetric slice at $t=0$ is $\sim \sqrt{N}$. This argument
can be generalized (up to a proportionality constant) to other geometries without knowing the functional form of the
eigenvalues.
\begin{equation}
	n_{\mx} = \mathrm{\con} {N}^{\frac{d-1}{d}}.
	\label{numtrunc}
\end{equation} 
In general, the Cauchy hypersurface can be deformed to minimize the spatial volume. Therefore, the choice of $\alpha$ is
neither unique nor covariant. In the examples below, we will see that the choice of this parameter is non-trivial and is
based on the coefficients we expect in the area law and on requiring complementarity.

Another possible truncation scheme, called ``linear truncation'' has been used. It is based on numerically identifying the location of the knee in the spectrum. This involves comparison of the slopes and detection of a rapid fall in the slopes in the spectrum. Again, the choice of what `rapid' means is captured by a single parameter $\delta$ and this choice is dictated by various factors as mentioned above. The advantage of this method is that it is independent of the geometry or any other detail about the QFT. The details of this scheme can be found in \cite{syx-entropy}.

Before proceeding we must ask what an area law looks like on a causal set. Since the areas in question are of  co-dimension $2$
surfaces, these are sets of measure zero in the causal set discretisation. On the other hand, given that there is a
length scale $\rho^{-1/d}$ associated with the discreteness scale, one can ascribe to the causal set a dimension
dependent scale  $N^{-1/d}$. Thus, we expect that an area law for $d>2$ should to be  of the form
 \begin{equation}
S^{cs}_{d} = a N^{\frac{d-2}{d}} + b. 
  \end{equation}  
For $d=2$ in the continuum the EE satisfies the log behaviour \cite{cc}
\begin{equation}
	\mathcal S= \frac{1}{3}\log\bigg(\frac{l}{\epsilon}\bigg)+b,
\end{equation}
where $\epsilon$ is the cut-off. Thus, in the causal set we expect that 
\begin{equation}
S^{cs}_{2} = a \ln N + b
  \end{equation}

The truncation procedure employed in \cite{soryaz} for the nested causal diamonds in $\mink^2$ and adapted to nested causal
diamonds in $\mink^4$ as well as $d=2,4$  de Sitter horizons can be summarised as below. 
\begin{equation}
	\begin{array}{ccccc} 
		i \hD &\text{truncation}&i \hD^{\text{t}}& \stackrel{SJ}{\Rightarrow} &\hW^{\text{t}}\\
		&&  \big\downarrow\  &  \text{restriction}  &\big\downarrow\\
		&&i \hD_{\cO}^{\text{t}}&  &\hW^{\text{t}}_{\cO}\\
		&&\big\downarrow&\text{truncation}&\big\downarrow\\
		&&i \hD^{\text{t}\text{t}}_{\cO}&&\hW^{\text{tt}}_{\cO}.
	\end{array} 
\end{equation} 
One starts  with the Pauli Jordan operator  $i \hD$ in $C$. Its spectrum is then  truncated to obtain $i
\hD^{\text{t}}$ and this gives a truncated  SJ Wightman function $\hW^{\text{t}}$. The restriction of
$\hW^{\text{t}}|_\cO$ to the subcausal set $\cO$ is however is not truncated with respect to the spectrum of $i
\hD$ in $\cO$. Hence  there is need for  a second  truncation of the SJ spectrum of  $i
\hD$ in $\cO$  which removes the large discrete UV contributions.  Thus, the double truncated $\hW^{\text{tt}}|_\cO$ is
used along with the truncated  $i \hD^{\text{t}}|_{\cO}$ to solve the generalised SSEE eigenvalue function in $\cO$.

\begin{figure}[!h]
	\begin{subfigure}{\textwidth}
		\includegraphics[width=0.3\textwidth]{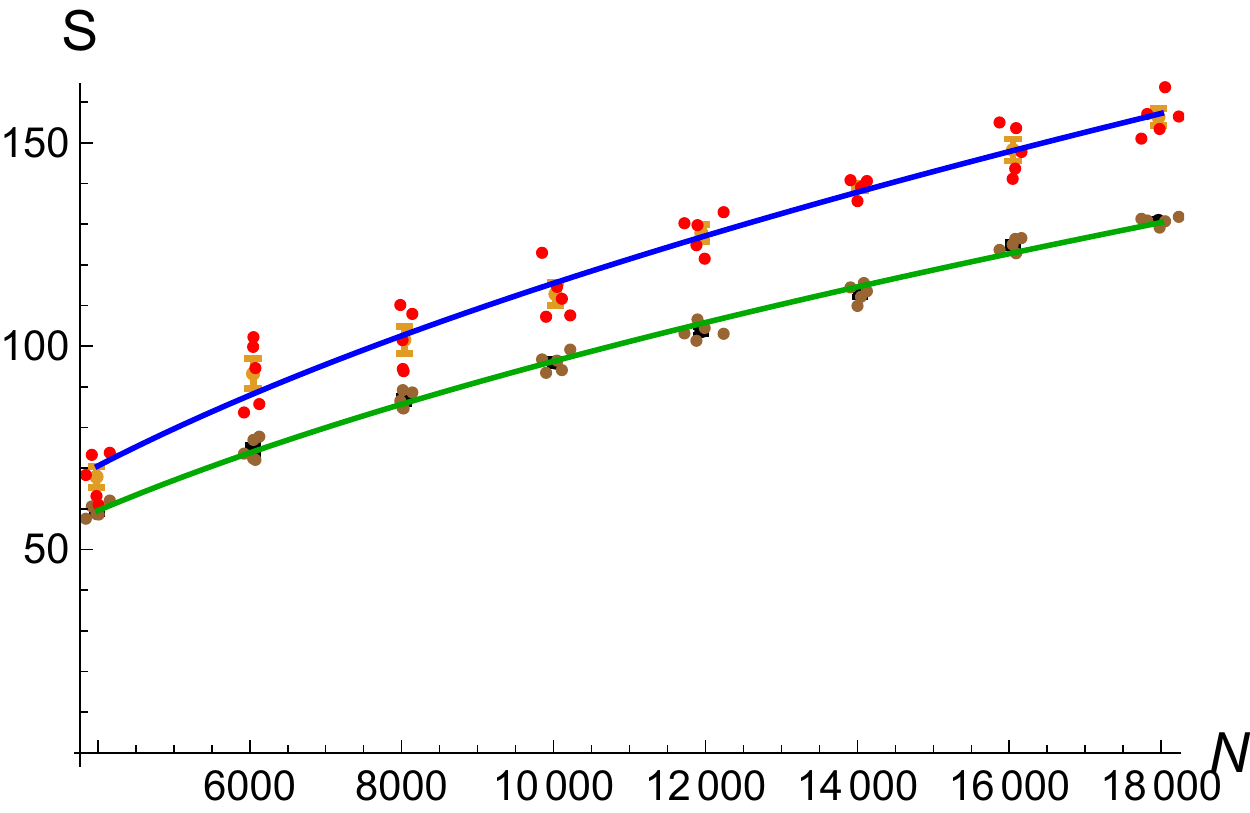}
		\includegraphics[width=0.3\textwidth]{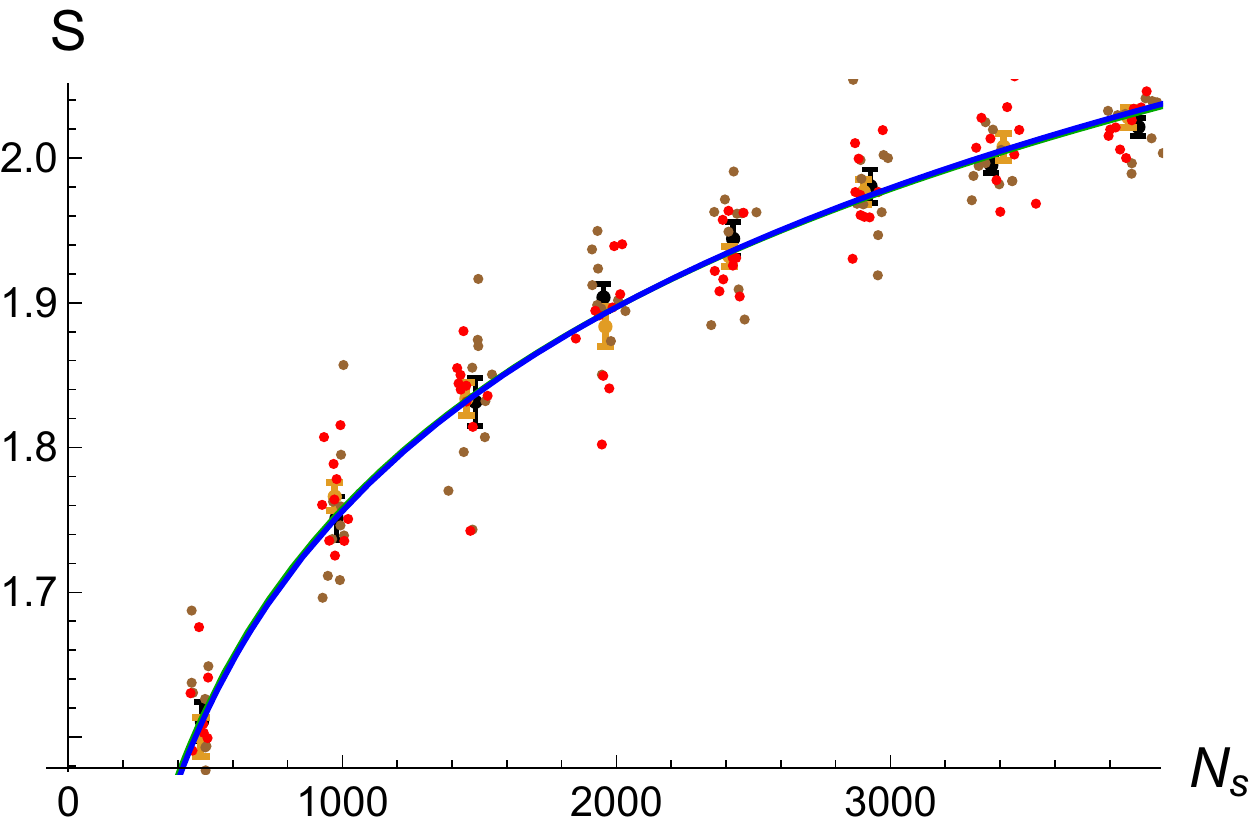}
		\includegraphics[width=0.3\textwidth]{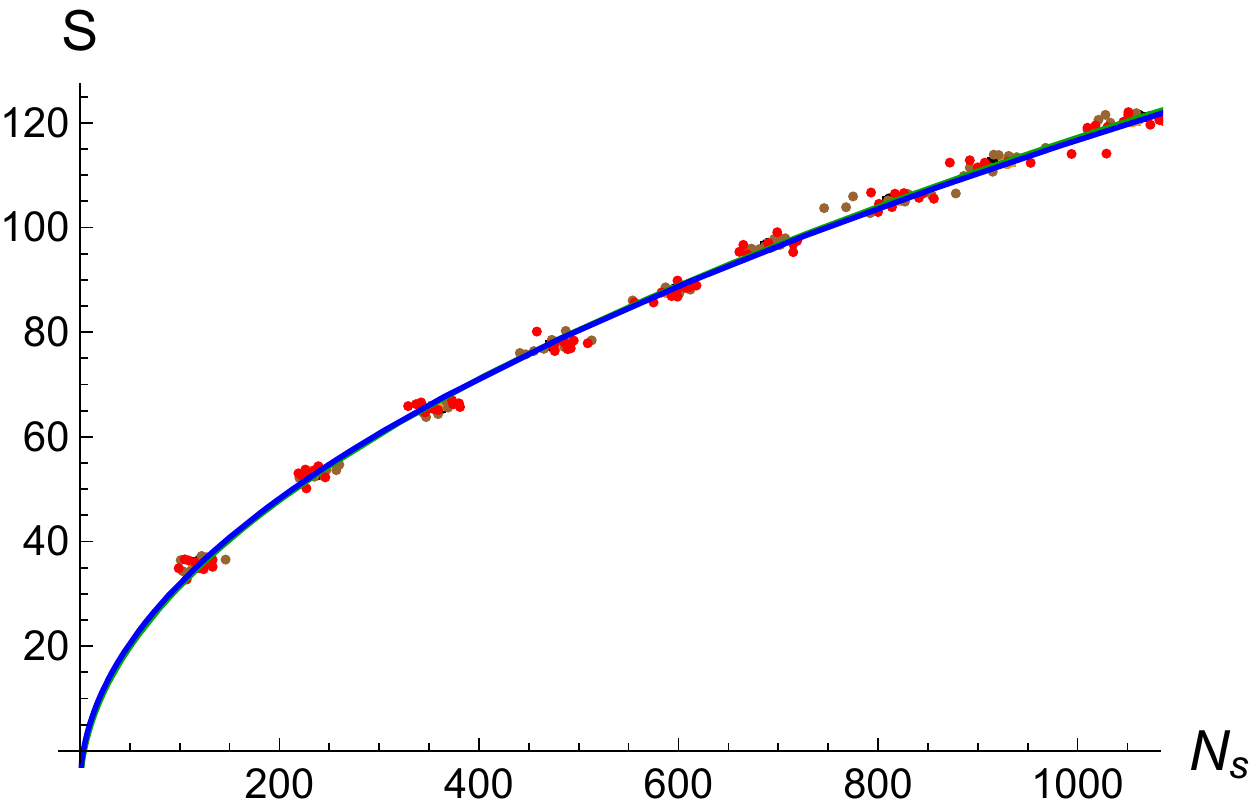}
		\caption{Number truncation} 
		\label{numtrunc}  
	\end{subfigure} 
	\begin{subfigure}{\textwidth}  
			\includegraphics[width=0.3\textwidth]{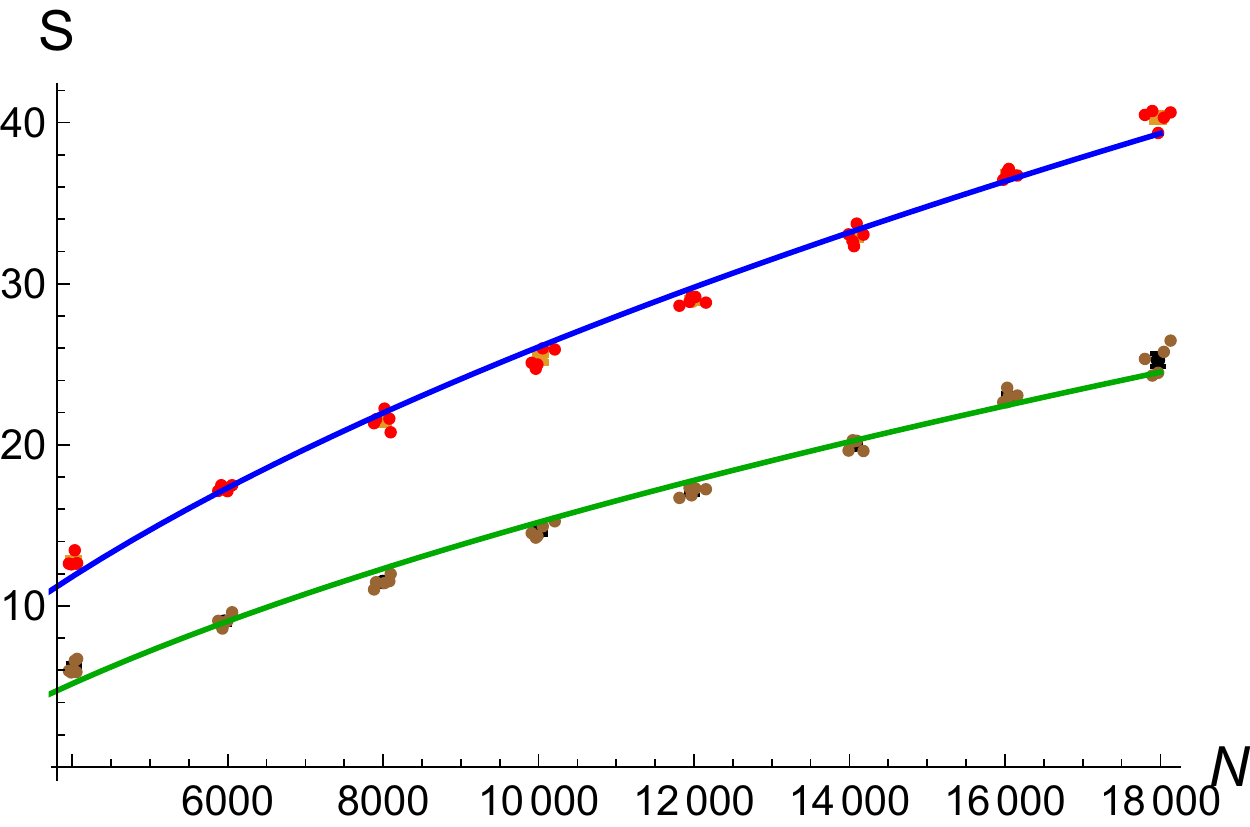}        
		\includegraphics[width=0.3\textwidth]{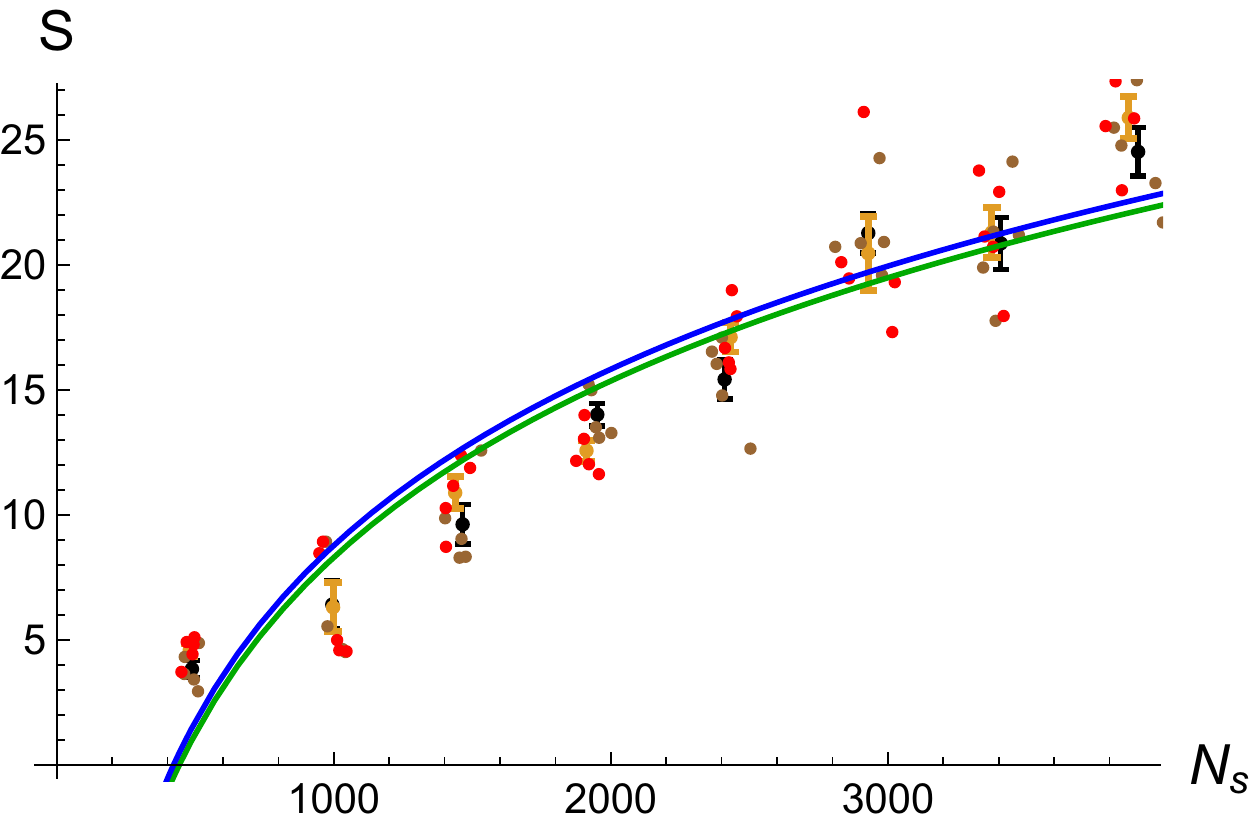}
		\includegraphics[width=0.3\textwidth]{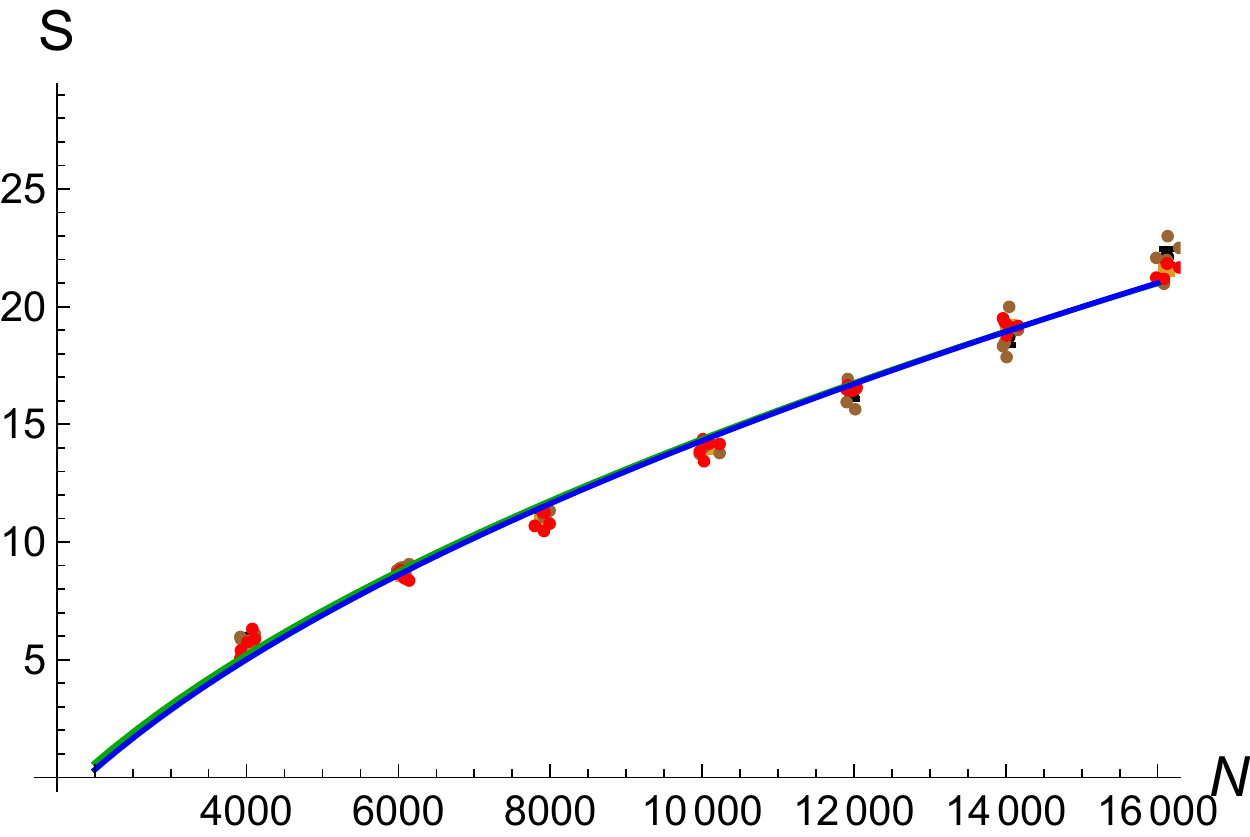}
		\caption{Linear truncation} 
		\label{lineartrunc}
	\end{subfigure}  
	\caption{SSEE vs. $N$ for the EE of horizons in $\mathbb{M}^4$ and de Sitter in 2 and 4 dimensions with two different truncations. Green and blue represent the data for the complementary regions.}
	\label{trunc_results}
\end{figure}

Figure \ref{trunc_results} summarises the results. What  is remarkable is that this truncation does what is expected of it -- it
restores an area law. The case of the nested causal diamonds in $\mink^4$ is however unsatisfying. The area law should
come hand in hand with complementarity, but this does not seem to be the case for any of the choices of truncation. In
the de Sitter case, we cannot check for complementarity since the de Sitter horizon is symmetric and the complementary
regions are equivalent. However, here too an area law emerges for suitable truncation schemes. 

\begin{figure}[h]\begin{center}
	\begin{subfigure}[b]{0.3\textwidth}
		\includegraphics[width=\textwidth]{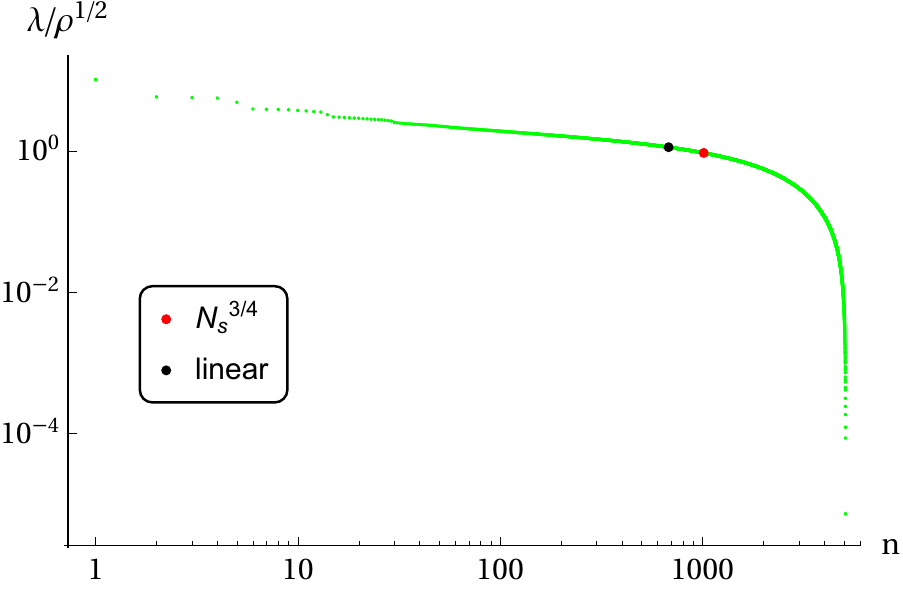}
		\caption{}
		\label{4dmink spectrum marked}
	\end{subfigure}
	\begin{subfigure}[b]{0.3\textwidth}
		\includegraphics[width=\textwidth]{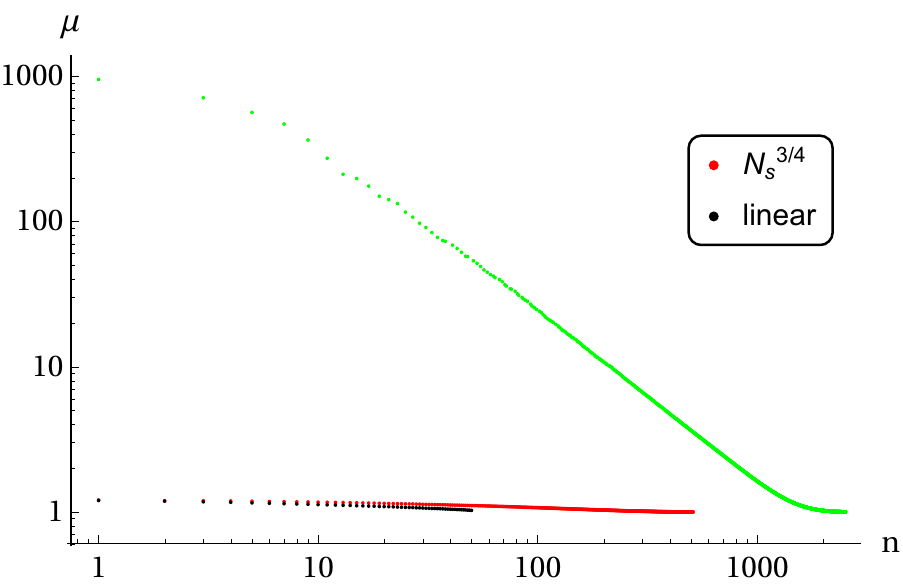}
		\caption{}
		\label{4dmink generalized spectrum}
	\end{subfigure}
	\caption{{\small For a causal diamond in $\mathbb{M}^4$ with $N=10k$, (a) is the spectrum of $i\Delta$ with different truncations marked, and (b) is a plot of the solutions of the generalised SSEE equation for these truncation schemes.}}
	\label{genspectra.fig}
        \end{center} 
\end{figure}

This   double truncation in the SJ spectrum  therefore has a profound effect  on the generalised
eigenvalue spectrum which defines the SSEE. In Fig \ref{genspectra.fig} this effect is shown for the nested causal
diamonds in $\mink^4$. What is remarkable is that the truncation leads to much smaller eigenvalues and hence
a much  smaller SSEE. In contrast, the untruncated generalised spectrum contains very large eigenvalues. 
% \section{Discussion}
% \label{disc}  

\section{Discussion}
\label{discussion}  

The foremost question that emerges is why the causal set SSEE has a volume dependence in the first place and how we
should interpret  the double truncation procedure since there {\it should} be no need for it in
an already finite theory. By trying to obtain a continuum-like result the worry is that one may be
unnecessarily throwing out an important  sign of new UV physics.  

As noted in \cite{X:2021dsa} the double truncation procedure leads to small violations of  causality. Namely, after  the
spectral truncation $\hD \rightarrow \hD^{\text{tt}}$,  $\Delta^{\text{tt}}(e,e')$ does not always vanish for spacelike
pairs $(e,e')$. These violations are ``small'' and become negligible when suitably averaged over the ensemble. Adding to
this is the work of  \cite{Keseman:2021dkf}  where the SJ  eigenfunctions in the $d=2$ 
causal diamond were examined in more detailed.  They found that the eigenfunctions beyond the knee  were highly
fluctuating at and ``below'' the discreteness scale.   They moreover vary considerably over the causal set 
ensemble unlike those in the scaling-regime which retain their general form.  
These results are an  indication that truncation may be physically justified in a
coarse grained, averaged sense, at length scales larger than  the truncation length scale. Indeed, in
\cite{soryaz,bbl} it has been suggested that since the SJ spectrum  beyond the knee contains modes with very small
eigenvalues, which are  
``nearly''  in $\ker \Delta$, the SSEE formula itself is not well defined.  Such modes are indeed what dominate the SSEE
in  the absence of 
truncation and give rise to  the volume law.

One conclusion is  that we cannot really speak of the QFT UV regime without  a full quantum theory
of causal sets. Indeed, in  all our discussions we have used only a classical causal set ensemble. It is plausible that  the
observables of the full theory are therefore such that one could still recovers an area law.

In the absence of such a full quantisation, however  it is reasonable to look for the effects  of causal
set discreteness  in a  phenomenologically interesting regime of quantum field theory. Manifold-like causal sets become
important when $\rho^{-1/d}$  is much larger than the Planck scale so that we  can talk of the continuum approximation,
while ignoring non-manifold-like contributions arising from full causal set quantum gravity \cite{kr,lc,mss,ccs}.  This  interim ``kinematic''  regime,  between known physics
and the Planck scale can lead to  interesting new physics as discussed in \cite{lambda,swerves}.  It is in this regime that we place
the above analysis of quantum field theory on causal sets. Rather than a ``complete'' quantum gravity theory of QFT we
wish to look at the regime in which discreteness does play a role.   Thus, far from being an artifact, the non-scaling UV behaviour of  the spectrum and the resultant volume law
for the SSEE could be a sign of new physics. 

It has for example 
been suggested in \cite{urbanapaper} that the effects of non-locality in quantum gravity could lead to volume laws for entanglement
entropy. If this were true, then we would need to understand the transition from a non-local (volume-law) regime to a local field theory (area-law) regime
as one moves away from the deep UV.  While the analysis  on causal sets shows this  explicitly with  the causal set  spectrum
exhibiting  a non-scaling behaviour in the
deep UV, it is important 
to understand this transition better, possibly from an RG perspective.  

An interesting question is whether there is a physical process which realises this transition to the deep
UV. Roughly, one might expect that ``entanglement probes'' with energies in the  scaling  regime would conclude that there is an area law, but those
that are more energetic, would  uncover a volume law. This
suggests a kind of ``screening'' of the interior of horizons for intermediate energy probes which gives rise to an {\it
  effective} area law, 
whereas  the horizon interiors are entangled with very high energy probes. 
Constructing an appropriate probe  in the causal set quantum field theory is of course a challenge. Given the emerging work 
 on volume laws in condensed matter systems it may be a fruitful first step  to construct 
suitable analogues in random lattice-like systems.

Of course there is the  bigger question of what a volume-law  might mean for blackhole evaporation, but given that such
systems are challenging to study in causal sets, we leave that as a  question for the future. 

\noindent {\bf Acknowledgements:} We would like to thank Yasaman Yazdi and Maximillian Ruep for discussions. NX is
supported by the AARMS fellowship at UNB.

\noindent {\sl Data sharing not applicable to this article as no datasets were generated or analysed during the current
  study.}

\bibliography{review_refs}
\bibliographystyle{ieeetr}
\end{document}